\documentclass[journal]{IEEEtran}
\usepackage{cite}
\ifCLASSINFOpdf
  \usepackage[pdftex]{graphicx}
  \graphicspath{{../pdf/}{../jpeg/}}
   \DeclareGraphicsExtensions{.pdf,.jpeg,.png}
\else
  \usepackage[dvips]{graphicx}
  \graphicspath{{../eps/}}
   \DeclareGraphicsExtensions{.eps}
\fi

\usepackage[cmex10]{amsmath}
\usepackage{array}
\ifCLASSOPTIONcompsoc
  \usepackage[caption=false,font=normalsize,labelfont=sf,textfont=sf]{subfig}
\else
 \usepackage[caption=false,font=footnotesize]{subfig}
\fi
\usepackage{amsmath,graphicx}
\usepackage{amssymb}
\usepackage{amscd}

\def\expk{{e^{-i \frac{2\pi}{T}k t}}}
\begin{document}

\title{FoCUS: Fourier-based Coded Ultrasound}

\author{Almog Lahav, 
Tanya Chernyakova
,~\IEEEmembership{Student Member,~IEEE}, Yonina C. Eldar,~\IEEEmembership{Fellow,~IEEE}}

\maketitle

\begin{abstract}
Modern imaging systems typically use single-carrier short pulses for transducer excitation.
Coded signals together with pulse compression are successfully used in radar and communication to increase the amount of transmitted energy.
Previous research verified significant improvement in SNR and imaging depth for ultrasound imaging with coded signals.
Since pulse compression needs to be applied at each transducer element, the implementation of coded excitation (CE) in array imaging is computationally complex. Applying pulse compression on the beamformer output reduces the computational load but also degrades both the axial and lateral point spread function (PSF) compromising image quality. In this work we present an approach for efficient implementation of pulse compression by integrating it into frequency domain beamforming. This method leads to significant reduction in the amount of computations without affecting axial resolution. The lateral resolution is dictated by the factor of savings in computational load.
We verify the performance of our method on a Verasonics imaging system and compare the resulting images to time-domain processing. We show that up to 77 fold reduction in computational complexity can be achieved in a typical imaging setups. The efficient implementation makes CE a feasible approach in array imaging paving the way to enhanced SNR as well as improved imaging depth and frame-rate.
\end{abstract}

\IEEEpeerreviewmaketitle

\section{Introduction}
\label{sec:intro}
Ultrasound is a radiation free imaging modality with numerous applications. 
In standard ultrasound systems an array of transducer elements transmits a short single-carrier Gaussian pulse. During its propagation echoes are scattered by acoustic impedance perturbations in the tissue, and received by the array elements. These echoes are essentially a stream of replicas of the transmitted pulse implying that the axial resolution is defined by the pulse duration. The data, collected by the transducers, is sampled and digitally integrated in a way referred to as beamforming, yielding a signal steered in a predefined direction and optimally focused at each depth. Such a beamformed signal, referred to as beam, forms a line in the image.

While the resolution is defined by the pulse duration, the resulting signal-to-noise ratio (SNR) and imaging depth are proportional to the amount of transmitted energy. One way to increase the energy is to transmit a longer pulse at a cost of resolution.
Increasing the energy while retaining the same pulse duration, requires higher peak intensity levels, which can potentially damage the tissue and thus are limited by the Food and Drug Administration (FDA) regulations.
More elaborate coded signals are used in radar and communication to overcome the above trade-off between the transmitted energy and resolution.

When coded signals are used for excitation, pulse compression is performed on the detected signal by applying a matched filter (MF) defined by the transmitted pulse-shape. As a result a stream of pulse's replicas is converted to a stream of its autocorrelations. The width of the pulse's autocorrelation is on the order of the inverse bandwidth \cite{rihaczek1969principles}, implying that the resolution is now defined only by the available system's bandwidth and is independent of pulse duration. Therefore, a longer pulse can be used to transmit more energy without degrading axial resolution. In coded ultrasound imaging phase-modulated signals are of interest since amplitude modulation is suboptimal in terms of energy.

Extensive studies show that, despite the frequency dependent attenuation characterizing biological tissues and the differences between detection and imaging nature of radar and ultrasound, coded signals can be successfully used in medical imaging \cite{chiao2005coded,misaridis2005use2}.
Experimental results reported in \cite{o1992coded} show improvement of $15$-$20$ dB in SNR as well as $30$-$40$ mm improvement in penetration depth. Due to these superior properties coded excitation (CE) was successfully applied in numerous ultrasound modalities including B-mode, flow and strain imaging \cite{liu2005coded,vogt2005development} as well as contrast imaging \cite{novell2009contrast,muzilla1999method}. 

Increased penetration and SNR are even more crucial for applications where the amount of transmitted energy is inherently reduced, e.g. synthetic aperture and plane wave imaging \cite{gammelmark2003multielement,o2005coded,song2015coded}. The feasibility of CE in plane-wave based shear wave motion detection was recently studied, showing substantial performance improvement \cite{song2015coded}.
Beyond the improvement in imaging depth, high SNR allows the utilization of higher frequencies and consequently better image resolution \cite{lewandowski2008high}. In addition, Misaridis and Jensen have shown a way to increase the frame rate by an orthogonal coding approach \cite{misaridis2005use1,misaridis2005use2,misaridis2005use3}.

\subsection{Challenges and Existing Solutions}
\label{ssec:challenges and solutions}
Despite the proven advantages of coded imaging its application in commercial medical systems is still very limited.
One of the main challenges of CE in medical ultrasound is its application to imaging with an array of transducer elements due to computational complexity of the required processing \cite{bjerngaard2002should,chiao2005coded,yoon2013efficient,ramalli2015real}. 
Regardless of the type of transmitted signal, sampling rates required to perform high resolution digital beamforming are typically significantly higher than the Nyquist rate of the signal. Rates up to $4$-$10$ times the central frequency of the transmitted pulse are used in order to eliminate artifacts caused by digital implementation of beamforming in time \cite{steinberg1992digital}. Taking into account the number of transducer elements the amount of sampled data that needs to be transferred to the processing unit and digitally processed in real time is very large. The pulse compression step, required in coded imaging, further increases the computational load.

Implementation of CE in array imaging requires a MF for every transducer element increasing the computational complexity.
Most reported experimental studies use either a single transducer element \cite{misaridis2005use2,novell2009contrast,vogt2005development} or an array of elements with one MF applied on the beamformed output \cite{o1992coded,chiao2005coded}. 
%
%
Even though the advantage of a single application of a MF on the beamformed data is obvious from the computational perspective, this approach degrades the performance of pulse compression \cite{chiao2005coded}.
The process of beamforming is comprised of averaging the received signals after their alignment with appropriate delays. To obtain dynamic focusing the applied delays are non-linear and time dependent and, thus, distort the phases of the coded signals. Due to this distortion the sidelobe level and the main lobe width of the MF output are degraded, leading to decreased contrast and resolution, which are highly prominent at low imaging depth.

One way to minimize this effect is to limit the duration of the transmitted pulse \cite{o1992coded}. However, this obviously reduces the amount of transmitted energy and, thus, goes against the main motivation behind the usage of coded signals.
Another approach is based on the results reported in \cite{bjerngaard2002should}. There it is shown that the degradation of the compression performance due to dynamic beamforming depends on the depth of the transmit focus. A possible solution is, therefore, to divide an image to several depth zones and adjust the code length according to the depth in order to reduce the compression error. In this case, however, the number of transmission events is increased in accordance to the number of focal zones, which in turn increases the acquisition time.

\subsection{Contributions}
\label{ssec:contributions}
Here we propose an approach that achieves perfect pulse compression, while keeping the computational complexity low. Our method is based on incorporating pulse compression into frequency domain beamforming (FDBF) developed recently for medical ultrasound \cite{wagner2012compressed,chernyakova2014compressed,burshtein2016sub}. This allows to use a MF in each transducer element at a low cost.

The concept of beamforming in frequency was first addressed back in the sixties for sonar arrays operating in the far field. However, translating these ideas to ultrasound imaging is much more involved due to near field operation, requiring non-linear beamforming. To the best of our knowledge, it was first addressed in \cite{wagner2012compressed,chernyakova2014compressed}, where it was shown that in the frequency domain the Fourier components of the beamformed signal can be computed as a weighted average of those of the individual detected signals. 
Since the beam is obtained directly in frequency, its Fourier components are computed only within its effective bandwidth. This is done using  generalized low-rate samples of the received signals, implying sampling and processing at the effective Nyquist rate which is defined with respect to the signals effective bandpass bandwidth \cite{eldar2015sampling}.

We note that according to the convolution theorem, pulse compression applied at each transducer element through MF is equivalent to multiplication in the frequency domain. In this work we show that it can be applied by appropriate modification of weights required for FDBF. As a result, not only is beamforming performed in frequency at a low rate, but it also includes MF of each individual channel without any additional computational load to the FDFB technique. The proposed method, performing both beamforming and pulse compression in frequency, is referred to as FoCUS, Fourier-based coded ultrasound.
The performance of our method is verified on a Verasonics ultrasound system and is compared to time-domain processing in terms of lateral and axial resolution. We evaluate the computational load for typical imaging setup and show that it can be reduced by 4 to 77 fold compared to time-domain implementation, depending on the oversampling factor and the required lateral resolution.
This efficient implementation allows CE to become a practical approach in array imaging.

The rest of the paper is organized as follows: Section \ref{sec:CEinUS} reviews basics of CE applied to medical imaging. In Section \ref{sec:ArrayImaging} we discuss the requirements and challenges of CE in the context of array imaging. We next propose a solution based on frequency domain beamforming in Section \ref{sec:BF and PC in frequncy}. The experimental results and the performance of FoCUS in terms of image quality and computational complexity are presented in Section \ref{sec:Results}.


\section{Coded Excitation in Medical Ultrasound}
\label{sec:CEinUS}
In CE a modulated signal is used for transducer excitation
\begin{equation}
\label{eq:mod signal}
s(t)=a(t)\cos(2\pi f_0t+\psi(t)),~~ 0\leq t\leq T_p.
\end{equation}
where $\psi(t)$ and $a(t)$ are phase and amplitude modulation functions respectively, $f_0$ is the central frequency of a transducer and $T_p$ is the duration of the signal. Assuming $L$ scatterers along the transmitted signal propagation path, the reflected signal, $\varphi(t)$, detected by an individual transducer element is given by
\begin{equation}
\label{eq:received signal}
\varphi(t)=\sum_{l=1}^L{\alpha_ls(t-t_l)},
\end{equation}
where $\{\alpha_l\}_{l=1}^L$ and $\{t_l\}_{l=1}^L$ are amplitudes and delays defined respectively by the scatterer's reflectivity and location.

Next, pulse compression is performed on the detected signal, $\varphi(t)$, by applying a MF, $h(t)=s^*(-t)$. The output is then a combination of autocorrelations of the transmitted pulse \cite{misaridis2005use1}
\begin{equation}
\label{eq:autocor}
\varphi^{CE}(t)=\varphi(t)\ast s^*(-t)=\sum_{l=1}^L{\alpha_lR_{ss}(t-t_l)}.
\end{equation}
The autocorrelation is given by
\begin{equation}
\label{eq:autocor_def}
R_{ss}(t)=\int_{-\infty}^{\infty}s(\tau)s^*(\tau-t)d\tau.
\end{equation}
The half-power width of the main lobe of the autocorrelation, which determines the range resolution, is approximately equal to the inverse bandwidth $B^{-1}$ \cite{rihaczek1969principles} of the transmitted pulse. As a result, in contrast to the conventional approach, the pulse time duration, $T_p$ , can be increased and more energy transmitted without degrading range resolution. The resulting gain in signal-to-noise ratio of the MF processing is approximately equal to the time-bandwidth product $D=T_pB$ \cite{wehner1995high}.

The above result holds when the detected signal is comprised of the exact replicas of the transmitted pulse. In practice, when acoustic wave propagates in biological tissues, high frequencies undergo stronger attenuation due to the medium's properties. A common way to model this effect is to assume that it does not distort the complex envelope of the reflected signal and only downshifts its central frequency \cite{jensen1996estimation}. The pulse reflected from the $l$th scatterer is then given by 
\begin{equation}
\label{eq:signal_freq_shift}
s(t-t_l,f_l)=a(t-t_l)\cos(2\pi (f_0-f_l)(t-t_l)+\psi(t-t_l)).
\end{equation}
As a result, the MF output depends on the central frequency shift, $f_l$, and is given by
\begin{equation}
\label{eq:ambiguity func_def}
A(t-t_l,f_l)=\int_{-\infty}^{\infty}s(\tau-t_l,f_l)s^*(\tau-t)d\tau.
\end{equation}
The function $A(t,f)$ is referred to as the ambiguity function.
Similar to \eqref{eq:autocor}, for $L$ scatterers the output of the MF is a stream of cross-sections of ambiguity function
%
\begin{equation}
\label{eq:ambiguity func}
\varphi^{CE}(t)=\varphi(t)\ast s^*(-t)=\sum_{l=1}^L{\alpha_lA(t-t_l,f_l)}.
\end{equation}

In ultrasound imaging the frequency shifts do not carry valuable information and thus do not need to be found explicitly. However, the width of the main lobe of the ambiguity function needs to be small for all values of $f_l$ to ensure good axial resolution for all frequency shifts.
This makes the ambiguity function of linear frequency modulation (FM) a good choice for ultrasound imaging.

The expression for a linear FM signal is
%
\begin{align}
\label{eq:chirp}
s(t)=a(t)\cos\left(2\pi\left((f_0-B/2)t+\frac{B}{2T_p}t^2\right)\right),~~ 0\leq t\leq T_p.
\end{align}
%
The frequency spectrum of the linear FM complex envelope is rectangular, so that the envelope of the MF output is approximately a sinc function \cite{cook1967radar}.
%
\begin{figure}[htb]
\begin{minipage}[b]{1\linewidth}
  \centering
  \centerline{\includegraphics[width=8cm]{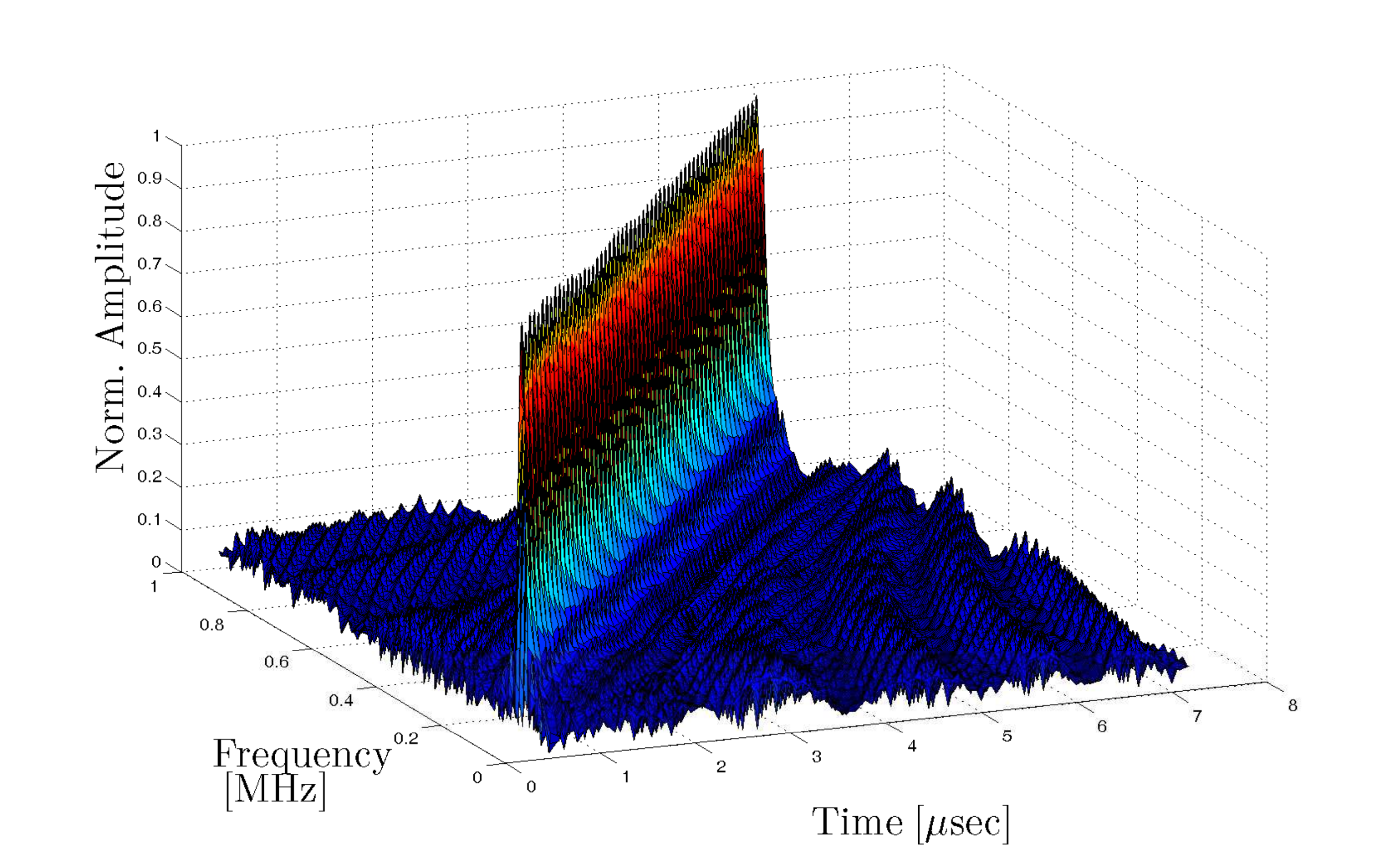}}
\end{minipage}
\caption{Ambiguity function of Linear FM with time-bandwidth product of $D=70$.}
\label{fig:ambiguityFunction}
\end{figure}
The ambiguity function of linear FM is shown in Fig. \ref{fig:ambiguityFunction}.
One can recognize the shape of a sinc in the cross sections parallel to the frequency axis.
Note that these cross sections preserve the same main lobe width for every frequency shift.

A significant improvement in penetration depth and contrast with linear FM excitation are reported in  \cite{misaridis2005use2}. However, these are obtained with a single-element transducer, while imaging systems today use an array of transducer elements. The implication of array processing is discussed next.

\section{Use of Coded Excitation in Array Imaging }
\label{sec:ArrayImaging}
\subsection{Conventional Array Imaging}
\label{ssec:DynamicBF}
Most commercial imaging systems today use multiple transducer elements to transmit and receive acoustic pulses. This allows for beamforming, a common signal-processing technique that enables spatial selectivity of signal transmission or reception \cite{van2004detection}.
In ultrasound imaging beamforming is used for steering the beam in a desired direction and focusing it in the region of interest in order to detect tissue structures.
Transmission beamforming, achieved by delaying the transmission time of each transducer element, allows for transmitting energy along a narrow beam. Beamforming upon reception is more involved. Here dynamically changing delays are applied on the signals received at each one of the transducer elements prior to averaging. Time-varying delays allow dynamic shift of the reception beam's focal point, optimizing angular resolution. Averaging of the delayed signals in turn enhances the SNR of the resulting beamformed signal, which is used to form a line in an image.
From here on, the term beamforming will refer to beamforming upon reception, which is the focus of this work.

\begin{figure}[htb]
\begin{minipage}[b]{1.0\linewidth}
  \centering
  \centerline{\includegraphics[width=6cm]{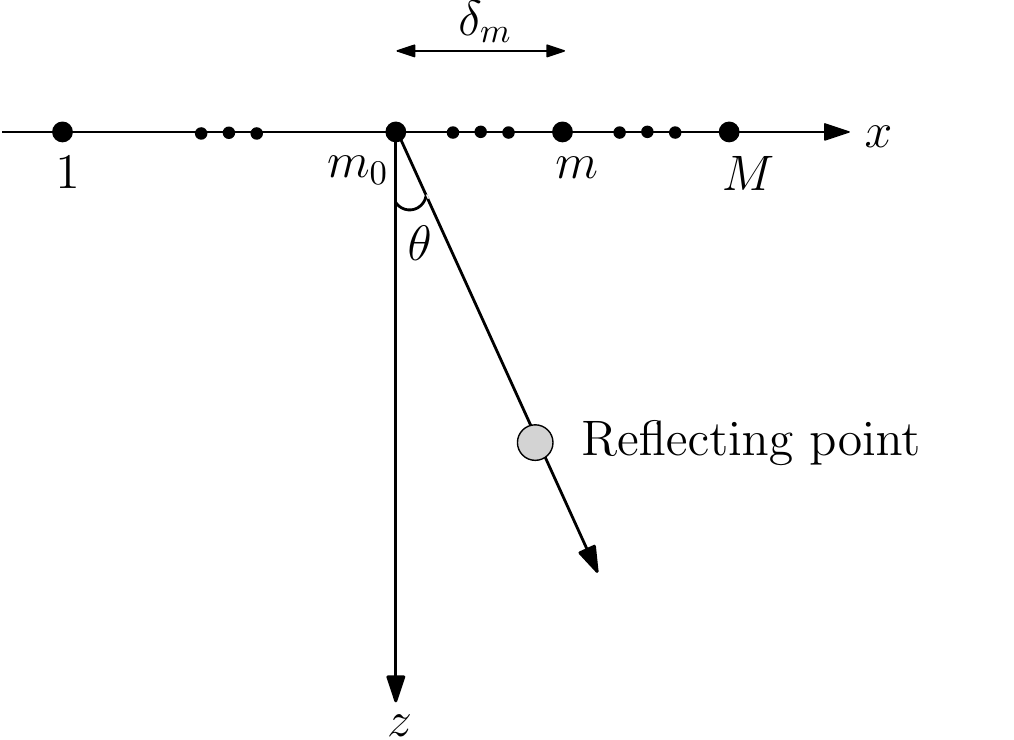}}
\end{minipage}
\caption{$M$ receivers aligned along the $x$ axis. An acoustic pulse is transmitted in a direction $\theta$.}
\label{fig:array}
\end{figure}

Consider an array of $M$ elements, illustrated in Fig. \ref{fig:array}. Denote by $m_0$ the reference element, by $\delta_m$ its distance to the $m$th element and by $c$ the speed of sound.
The image cycle begins at $t=0$, when the array transmits an energy pulse in the direction $\theta$. The pulse propagates within the tissue at speed $c$, and at time $t\geq0$ it reaches a potential point reflector located at $(x,z)=(ct\sin{\theta},ct\cos{\theta})$. The resulting echo is received by all array elements at a time defined by their locations. Denote by $\varphi_m(t)$ the signal received by the $m$th element and by $\hat{\tau}_m(t;\theta)$ the time of arrival. It is readily seen that
\begin{equation}
\label{eq:tau^m}
\hat{\tau}_m(t;\theta)=t+\frac{d_m(t;\theta)}{c},
\end{equation}
where $d_m(t;\theta)=\sqrt{(ct\cos{\theta})^2+(\delta_m-ct\sin{\theta)}^2}$ is the distance traveled by the reflection.
Beamforming involves averaging the signals received by multiple receivers while compensating for the differences in arrival time.

Since $\delta_{m_0}=0$, the arrival time at $m_0$ is $\hat{\tau}_{m_0}(t;\theta)=2t$. Applying an appropriate delay to $\varphi_m(t)$, such that the resulting signal $\hat{\varphi}_m(t;\theta)$ satisfies $\hat{\varphi}_m(2t;\theta)=\varphi_m(\hat{\tau}_m(t;\theta))$, we align the reflection received by the $m$th receiver with the one received at $m_0$. Denoting $\tau_m(t;\theta)=\hat{\tau}_m(t/2;\theta)$  and using \eqref{eq:tau^m}, the following aligned signal is obtained
\begin{align}
\label{eq:phim}
\hat{\varphi}_m(t;\theta)&=\varphi_m(\tau_m(t;\theta)),\\ \nonumber
\tau_m(t;\theta)&=\frac{1}{2}\left(t+\sqrt{t^2-4(\delta_m/c)t\sin{\theta}+4(\delta_m/c)^2}\right).
\end{align}
The beamformed signal may now be derived by averaging the aligned signals
\begin{equation}
\label{eq:phi beamformed}
\Phi(t;\theta)=\frac{1}{M}\sum_{m=1}^M{\hat{\varphi}_m(t;\theta)}=\frac{1}{M}\sum_{m=1}^M\varphi_m(\tau_m(t;\theta)).
\end{equation}
Such a beam is optimally focused at each depth and hence exhibits improved angular localization and enhanced SNR.

In practice, the processing defined in \eqref{eq:phi beamformed} is performed digitally, where the values of the detected signal $\varphi_m(t)$ at $\tau_m(t;\theta)$ are obtained by linear interpolation.
\subsection{Matched Filtering and Beamforming}
\label{ssec:MFandBF}

\begin{figure*}[htb]
\begin{minipage}[b]{1.0\linewidth}
  \centering
  \centerline{\includegraphics[width=12cm]{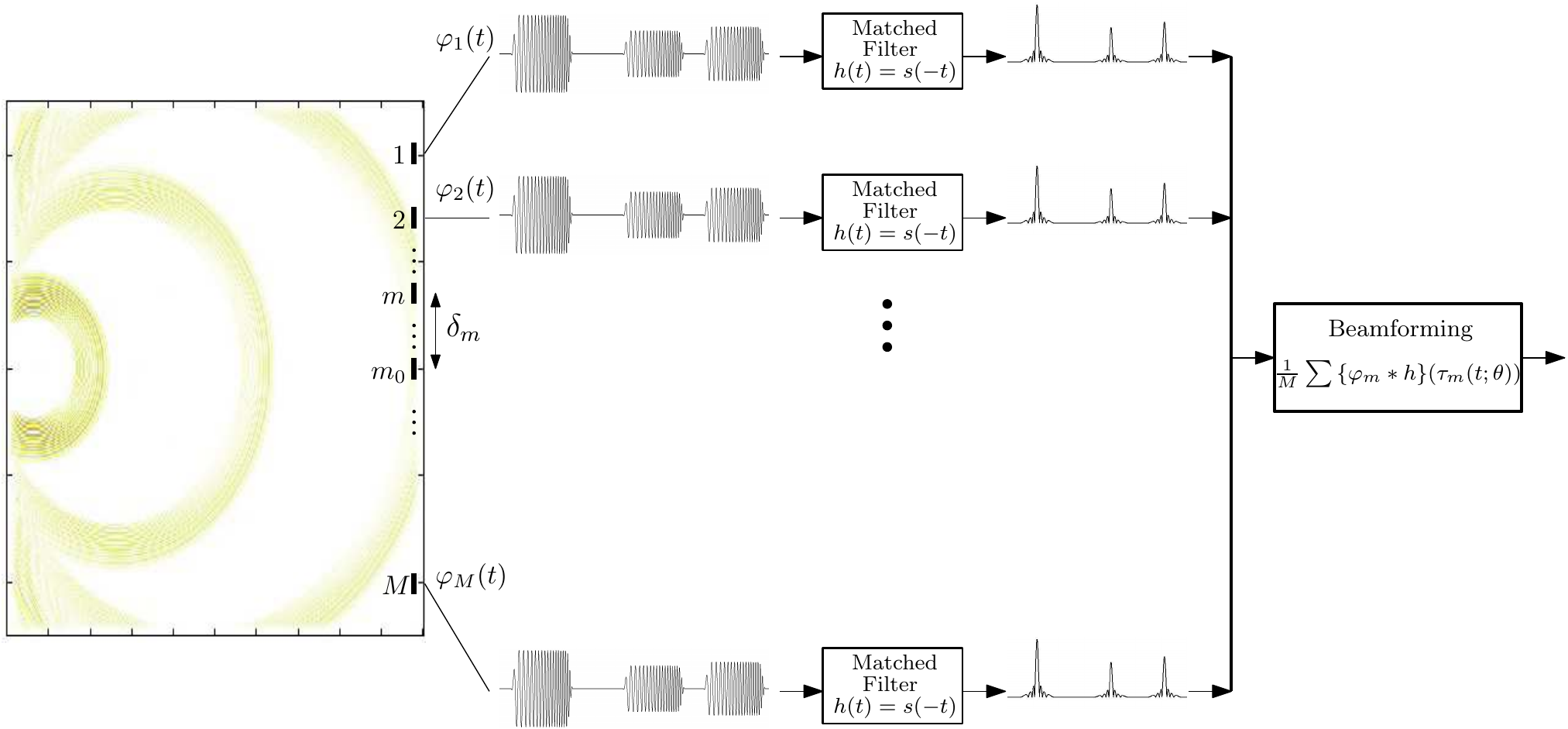}}
\end{minipage}
\caption{On the left echoes are reflected from 3 point scatterers in the medium. The received signal at each transducer element is composed of reflections of a linear FM waveform. The beamforming is applied on compressed signals, obtained at the output of the MF at each element.}
\label{fig:bf illustration}
\end{figure*}

As explained in Section \ref{sec:CEinUS}, in the CE approach, pulse compression is achieved by applying a MF on the detected signal. Array imaging requires matched filtering of the detected signal at every transducer element prior to beamforming as illustrated in Fig. \ref{fig:bf illustration}. This implementation is referred to as beamforming pre-compression \cite{bjerngaard2002should}, since beamforming is applied after compression. Using \eqref{eq:phi beamformed} and substituting the MF impulse response $h(t)=s^*(-t)$, the beamformed signal is given by
%
\begin{align}
\label{eq:phi beamformed CE}
\Phi_{CE}(t;\theta)&=\frac{1}{M}\sum_{m=1}^M{\varphi_m^{CE}(\tau_m(t;\theta))},\\ \nonumber
\varphi_m^{CE}(t)&=\{\varphi_m\ast h\}(t).
\end{align}
A practical meaning of this implementation is that the computational complexity is vastly increased by filtering each detected signal. This restricts the use of CE in array imaging.

A straightforward way to overcome this problem is beamforming post-compression, i.e. to perform beamforming using the uncompressed detected signals and apply MF on the beamformer's output \cite{bjerngaard2002should}.
This requires only one MF allowing to save:
\begin{equation}
\label{eq:N mul saved}
N\approx(M-1)\left(\frac{3}{2}(N_s+N_h)\log(N_s+N_h)+N_s+N_h\right),
\end{equation}
multiplications, where $M$, $N_s$ and $N_h$ are the number of elements, number of samples and MF length respectively.

The advantage of a single application of MF on the beamformed data is obvious from the computational perspective. However with this approach the resulting beamformed signal is given by
\begin{align}
\label{eq:phi beamformed CE post}
\Phi_{CE_{post}}(t;\theta)=\left\{\frac{1}{M}\sum_{m=1}^M{\varphi_m(\tau_m(t;\theta))}\right\}\ast h(t).
\end{align}
As can be seen in \eqref{eq:phim}, the applied delay is a non-linear function of time and varies within the support of the coded pulse. As a result the detected signal is distorted and is no longer comprised of exact replicas of the transmitted pulse. Due to this distortion the impulse response of the MF is mismatched with the input signal. This results in poor compression performance:
sidelobe level and main lobe width are degraded leading to decreased resolution and contrast. A detailed study of the effect of beamforming post-compression on image quality is presented in \cite{o1992coded,bjerngaard2002should}.

Figure \ref{fig:TDBF Vs FDBF single} shows scan-lines with a point scatterer located at 10 mm from the transducer obtained with beamforming post- and pre-compression. Details on the experiment environment are elaborated in Section \ref{sec:Results}. The main lobe of the resulting axial point spread function is approximately $13\%$ wider when beamforming is performed prior to pulse compression. In addition, the sidelobe is $9$ dB higher. This degradation can be easily seen in the resulting images, presented in Figs. \ref{fullImagePostComp} and \ref{fullImagePreComp}, corresponding to beamforming pre- and post-compression respectively. A zoom in on the three first point reflectors is shown in Figs. \ref{fullImageZoomPostComp} and \ref{fullImageZoomPreComp}.
When the trivial solution of beamforming post-compression is used in order to reduce the computational complexity the image quality is compromised.



As elaborated in Section \ref{ssec:challenges and solutions}, the existing approaches to minimize this effect either reduce the transmitted energy or increase acquisition time.
Obviously, efficient implementation of MF at each transducer channel prior to beamforming is a preferable approach for coded array imaging.

\section{Beamforming and Pulse Compression in the Frequency Domain}
\label{sec:BF and PC in frequncy}
%
To obtain a computationally efficient method to apply MF at each transducer element prior to beamforming, we propose integrating the MF into the FDBF framework.

As mentioned in Section \ref{sec:intro}, beamforming in frequency was first considered in the context of sonar array processing \cite{williams1968fast,rudnick1969digital}, where due to far field operation mode it corresponds to averaging the signals after applying constant delays. This process can be transferred to the frequency domain in a straightforward manner through the well-known time shifting property of the Fourier transform.
In the context of ultrasound imaging due to dynamic nature of the beamforming that implies the non-linearity and time dependence of the required delays, the translation to frequency is much more involved. A frequency domain formulation of beamforming was introduced in \cite{chernyakova2014compressed} and \cite{burshtein2016sub} for 2D imaging and 3D imaging respectively leading to significant reduction in sampling and processing rate.
We first review the above framework based on \cite{chernyakova2014compressed} and then show how to incorporate pulse compression into it without increasing the computational cost.

\subsection{Prior Art: Frequency Domain Beamforming}
\label{ssec:Prior Art: FDBF}
Ultrasound operates in extreme near field. Therefore, to obtain the far-field beampattern of the array allowing for spatial selectivity, dynamic focusing is required. The focal point is moved throughout the scan depth by applying time dependent delays $\tau_m(t;\theta)$ defined in \eqref{eq:phim}. Despite the non-linearity of the delays, the Fourier components of the beamformed signal can be computed as a weighted average of those of the individual detected signals \cite{chernyakova2014compressed}.
To this end denote the Fourier coefficients of $\Phi(t;\theta)$ with respect to the interval $T$, defined by the maximal scan depth, by
\begin{equation}
\label{eq:fourier coeff of beamformed 1}
c[k]=\frac{1}{T}\int_0^T I_{[0,T_B(\theta))}(t) \Phi(t;\theta)\expk dt,
\end{equation}
where $I_{[a,b)}$ is the indicator function equal to $1$ when $a\leq t<b$ and $0$~otherwise and $T_B(\theta)=\min_{1\leq m \leq M}{\tau_m^{-1}(T;\theta)}$ \cite{wagner2012compressed}.
Substituting \eqref{eq:phi beamformed} into \eqref{eq:fourier coeff of beamformed 1} results in
\begin{equation}
\label{eq:fourier coeff of beamformed 2}
c[k]=\frac{1}{M}\sum_{m=1}^M \hat{c}_m[k],
\end{equation}
where
\begin{align}
\label{eq:c_k_m}
\hat{c}_m[k]&=\frac{1}{T}\int_0^T I_{[0,T_B(\theta))}(t) \varphi_m(\tau_m(t;\theta))\expk dt\\ \nonumber
&=\frac{1}{T}\int_0^T \varphi_m(t)q_{k,m}(t;\theta)\expk dt.
\end{align}
The last equation stems from the variable substitution ~~~~~~~~~~$x=\tau_m(t;\theta)$,~required to obtain $c[k]$ as a function of non-delayed receive signals $\varphi_m(t)$. The delays are effectively applied through the so-called distortion function, $q_{k,m}(t;\theta)$ given by
\begin{align}\label{eq:q(t)}
q_{k,m}(t;\theta)=& I_{[|\gamma_m|,\tau_m(T;\theta))}(t) \left(1+\frac{\gamma_m^2\cos^2{\theta}}{(t-\gamma_m\sin{\theta})^2}\right)\times  \\*
&\exp{\left\{i\frac{2\pi}{T}k\frac{\gamma_m-t\sin{\theta}}{t-\gamma_m\sin{\theta}}\gamma_m\right\}}, \nonumber
\end{align}
with $\gamma_m=\delta_m/c$.

To derive a relationship between the Fourier coefficients of the beam and those of the received signals, we next replace $\varphi_m(t)$ by its Fourier coefficients $c_m[n]$ and rewrite \eqref{eq:c_k_m} as
\begin{align}\label{c k m fourier}
 \hat{c}_m[k]&=\sum_n c_m[n]\frac{1}{T}\int_0^T q_{k,m}(t;\theta){e^{-i \frac{2\pi}{T}(k-n) t}} dt \\
 &=\sum_n c_m[k-n]Q_{k,m;\theta}[n]. \nonumber
\end{align}
Here $Q_{k,m;\theta}[n]$, referred to as Q-coefficients, are the Fourier coefficients of the distortion function with respect to $[0,T)$. When substituted by its Fourier coefficients, the distortion function effectively transfers the beamforming delays defined in \eqref{eq:phim} to the frequency domain. The function $q_{k,m}(t;\theta)$ depends only on the array geometry and is independent of the received signals. Therefore, its Fourier coefficients can be computed off-line and used as a look-up-table (LUT) during the imaging cycle.
According to Proposition 1 in \cite{wagner2012compressed}, $\hat{c}_m[k]$ can be approximated sufficiently well with a finite number $N_q$ of Q-coefficients
\begin{equation}\label{eq:c k m N1N2}
\hat{c}_m[k]\simeq\sum_{n=-N_1}^{N_2}c_m[k-n]Q_{k,m;\theta}[n],
\end{equation}
where $N_q=N_2-N_1+1$. The choice of $N_q$ controls the approximation quality.
As reported in \cite{chernyakova2014compressed} for $n<-N_1$ and $n>N_2$ $\{Q_{k,m;\theta}[n]\}$ are several orders of magnitude lower and therefore can be neglected allowing for efficient implementation of beamforming in frequency. The choice of $N_q$ and its effect on image quality are discussed in detail in Section \ref{ssec:Imaging results and performance}.

Finally, substitution of \eqref{eq:c k m N1N2} into \eqref{eq:fourier coeff of beamformed 2} yields the desired relationship between the Fourier coefficients of the beam and the individual signals
\begin{equation}\label{eq:FourierDomainRelationship}
c[k]\simeq\frac{1}{M}\sum_{m=1}^M \sum_{n=-N_1}^{N_2}c_m[k-n]Q_{k,m;\theta}[n].
\end{equation}
Applying an inverse Fourier transform on $\{c[k]\}$ results in the beamformed signal in time. We then proceed to standard image generation steps which include log-compression and interpolation.


\subsection{Integrating Pulse Compression in Frequency Domain Beamforming}
\label{ssec:pulse compression in FDBF}
To incorporate pulse compression into the FDBF framework we aim to express the Fourier coefficients of the signal $\Phi _{CE}(t;\theta)$, obtained by beamforming the compressed signals, as a function of the Fourier coefficients of the individual detected signals. 
This will allow us to perform the compression by means of weighting in the frequency domain together with beamforming.

Following the steps in Section \ref{ssec:Prior Art: FDBF} for $\Phi _{CE}(t;\theta)$, its Fourier coefficients  are given by
\begin{align}
\label{eq:fourier coeff of beamformed 2 CE}
c_{CE}[k]&=\frac{1}{M}\sum_{m=1}^M \hat{c}_m^{CE}[k], \\ \nonumber
\hat{c}_m^{CE}[k]=\frac{1}{T}\int_0^T &\varphi_m^{CE}(t)q_{k,m}(t;\theta)\expk dt,
\end{align}
with $q_{k,m}(t;\theta)$ given in \eqref{eq:q(t)}.
The next step is to replace $\varphi_m^{CE}(t)$ by its Fourier coefficients $c_m^{CE}[n]$
\begin{align}\label{eq: c k m N1N2 MF 1}
\hat{c}_m^{CE}[k]&\simeq\sum_{n=-N_1}^{N_2}c_m^{CE}[k-n]Q_{k,m;\theta}[n].
\end{align}
According to \eqref{eq:phi beamformed CE} and the convolution theorem, $c_m^{CE}[n]$ are given by $c_m[n] h[n]$, where $c_m[n]$ and $h[n]$ are the Fourier coefficients of the signal $\varphi_m(t)$ and the MF respectively. This allows us to rewrite \eqref{eq:c k m N1N2} as follows
\begin{align}\label{eq: c k m N1N2 MF}
\hat{c}_m^{CE}[k]&\simeq\sum_{n=-N_1}^{N_2}c_m[k-n]h[k-n]Q_{k,m;\theta}[n] \\
&=\sum_{n=-N_1}^{N_2}c_m[k-n]\tilde{Q}_{k,m;\theta}[n] \nonumber
\end{align}

\begin{figure}
\begin{minipage}[b]{1.0\linewidth}
  \centering
  \centerline{\includegraphics[width=7cm]{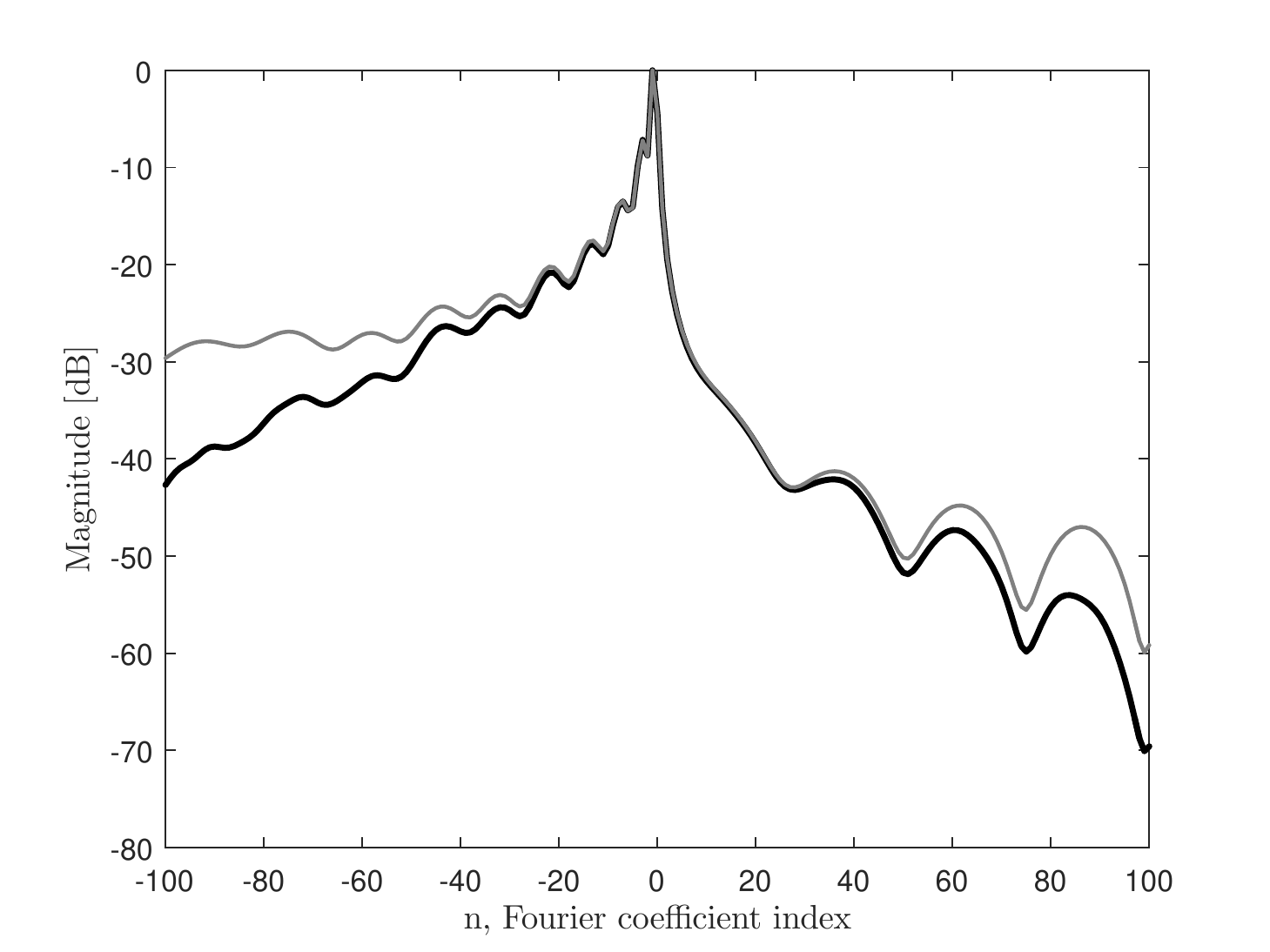}}
\end{minipage}
\caption{Fourier coefficients  $\tilde{Q}_{k,m;\theta}[n]$ and ${Q}_{k,m;\theta}[n]$ are characterized by a rapid decay.  In black bold line $\tilde{Q}_{k,m;\theta}[n]$. In gray thin line ${Q}_{k,m;\theta}[n]$. ($k=130$, element number $m=14$, angle $\theta=0.427 [\textrm{rad}]$).}
\label{fig:Qdecay}
\end{figure}

The decaying property of $\{Q_{k,m;\theta}[n]\}$ is retained after integration of MF: numerical studies show that most of the energy of $\{\tilde{Q}_{k,m;\theta}[n]\}$ is
concentrated around the DC component, irrespective of the choice of $k,m$ and $\theta$. We display these decay properties in Fig. \ref{fig:Qdecay}, where $\{\tilde{Q}_{k,m;\theta}[n]\}$ is plotted as a function of $n$ for $k=130$, $m=14$, $\theta=0.427~[\textrm{rad}]$.
%

Obviously, incorporation of pulse compression does not affect computational complexity of frequency domain beamforming, since it only requires to update the set of frequency weights which is performed off-line.
Substitution of \eqref{eq: c k m N1N2 MF} into \eqref{eq:fourier coeff of beamformed 2 CE} yields a relationship between the Fourier coefficients of the beam and the individual detected signals:
\begin{equation}\label{eq:FourierDomainRelationship CE}
c_{CE}[k]\simeq\frac{1}{M}\sum_{m=1}^M \sum_{n=-N_1}^{N_2}c_m[k-n]\tilde{Q}_{k,m;\theta}[n].
\end{equation}
Applying an inverse Fourier transform on $\{c_{CE}[k]\}$ results in time-domain output of beamforming the detected compressed signals.
%

\subsection{Processing at the Effective Nyquist Rate}
\label{ssec:processing at the effective Nyquist rate }

We next define the number of Fourier coefficients of the beamformed signal $\Phi_{CE}(t;\theta)$ that need to be computed using \eqref{eq:FourierDomainRelationship CE} and explain how the Fourier coefficients of the individual detected signals, required for computation, are obtained.
Denote by $\gamma$ the set of Fourier coefficients of the MF output, $\varphi_m^{CE}(t)$, that corresponds to its bandwidth, namely, the values of $k$ for which $c_m^{CE}[k]$ are nonzero (or larger than a threshold). Let $G$ denote the cardinality of $\gamma$.
Note that according to \eqref{eq:fourier coeff of beamformed 2 CE} and \eqref{eq: c k m N1N2 MF 1} the bandwidth of the beamformed signal will contain at most $K=G+N_q$ nonzero frequency components. In a typical imaging setup the value of $G$ is on the order of hundreds, while $N_q$, defined by the decay properties of $\{Q_{k,m;\theta}[n]\}$, are typically on the order of tens. This implies that $K\approx G$, so that the bandwidth of the beam is approximately equal to the bandwidth of the MF output.

According to \eqref{eq:FourierDomainRelationship CE}, $K$ Fourier coefficients of the beamformed signal are computed using at most $K+N_q$ Fourier coefficients of individual detected signals.
The above can be obtained from $K+N_q$ point-wise samples of the detected signal filtered with an analog kernel $s^*(t)$. In ultrasound imaging with modulated pulses the transmitted signal has one main band of energy. As a result the analog filter takes on the form of a band-pass filter, leading to a simple low-rate sampling scheme \cite{chernyakova2014compressed}.

The computational complexity of FoCUS and the achieved savings are discussed in detail in Section \ref{ssec:comp complexity}.
%
\begin{figure}[!h]
\begin{minipage}[b]{1.0\linewidth}
  \centering
  \centerline{\includegraphics[width=7cm]{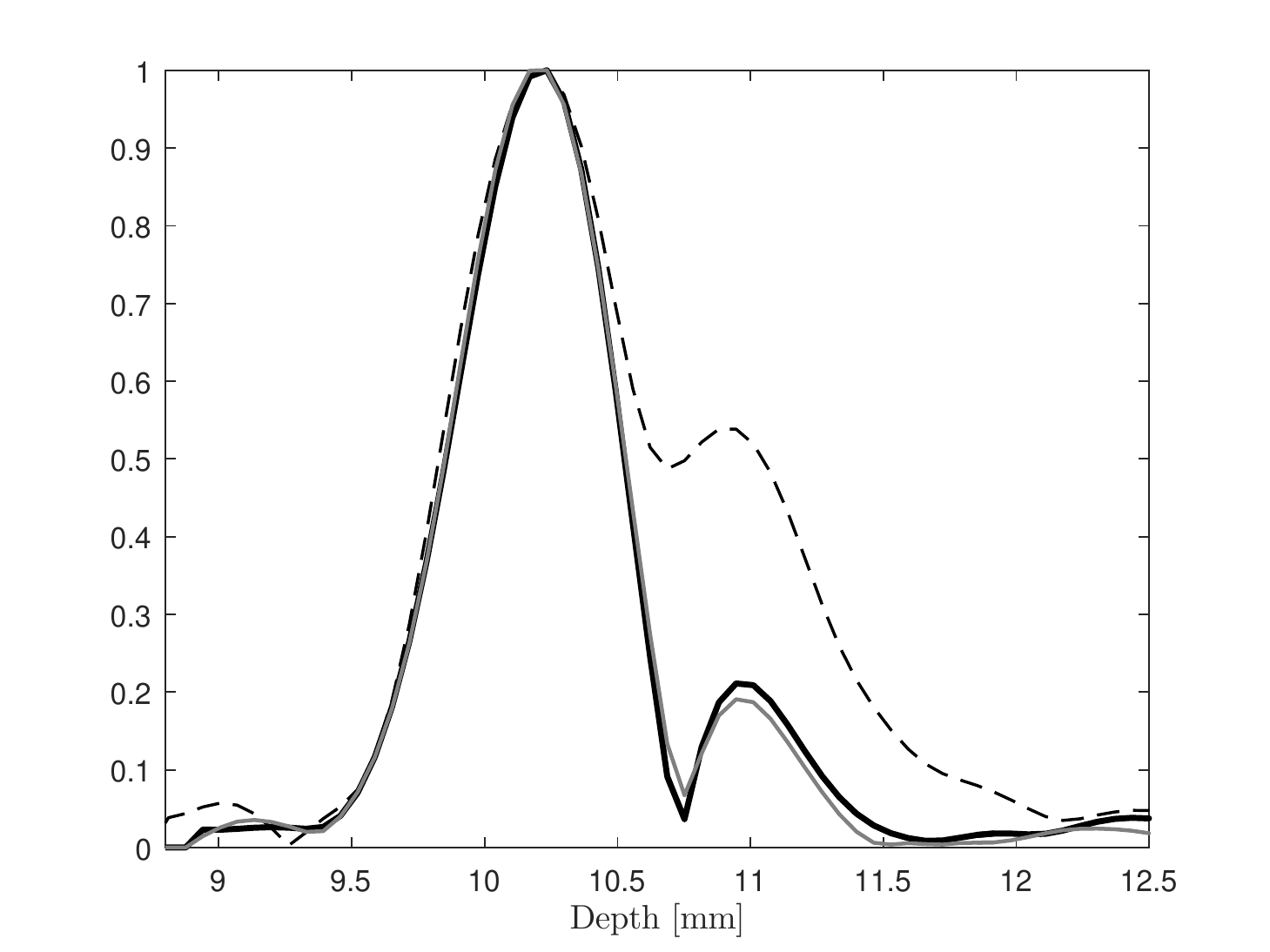}}
\vspace{-0.3cm}
\end{minipage}
\caption{A single scan-line of point scatterer positioned at 10 mm. Frequency domain processing in black bold line, beamforming pre-compression in thin gray line and beamforming post-compression in dashed line.}
\label{fig:TDBF Vs FDBF single}
\end{figure}

\section{Results}
\label{sec:Results}
\begin{figure*}[!t]
\centering
\subfloat[]{\includegraphics[width=9 cm]{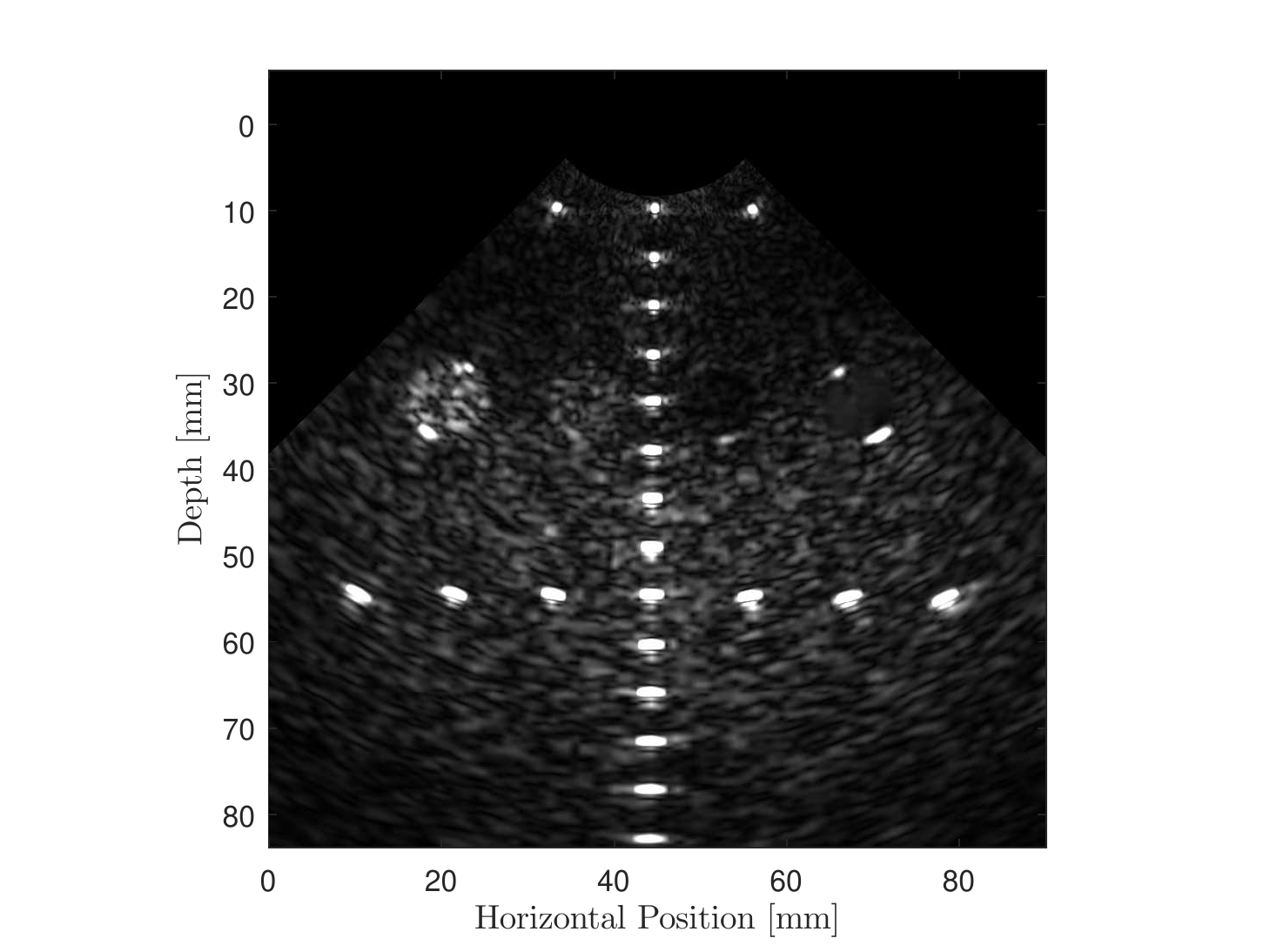}%
\label{fullImagePostComp}}
\hfil
\subfloat[]{\includegraphics[width=9 cm]{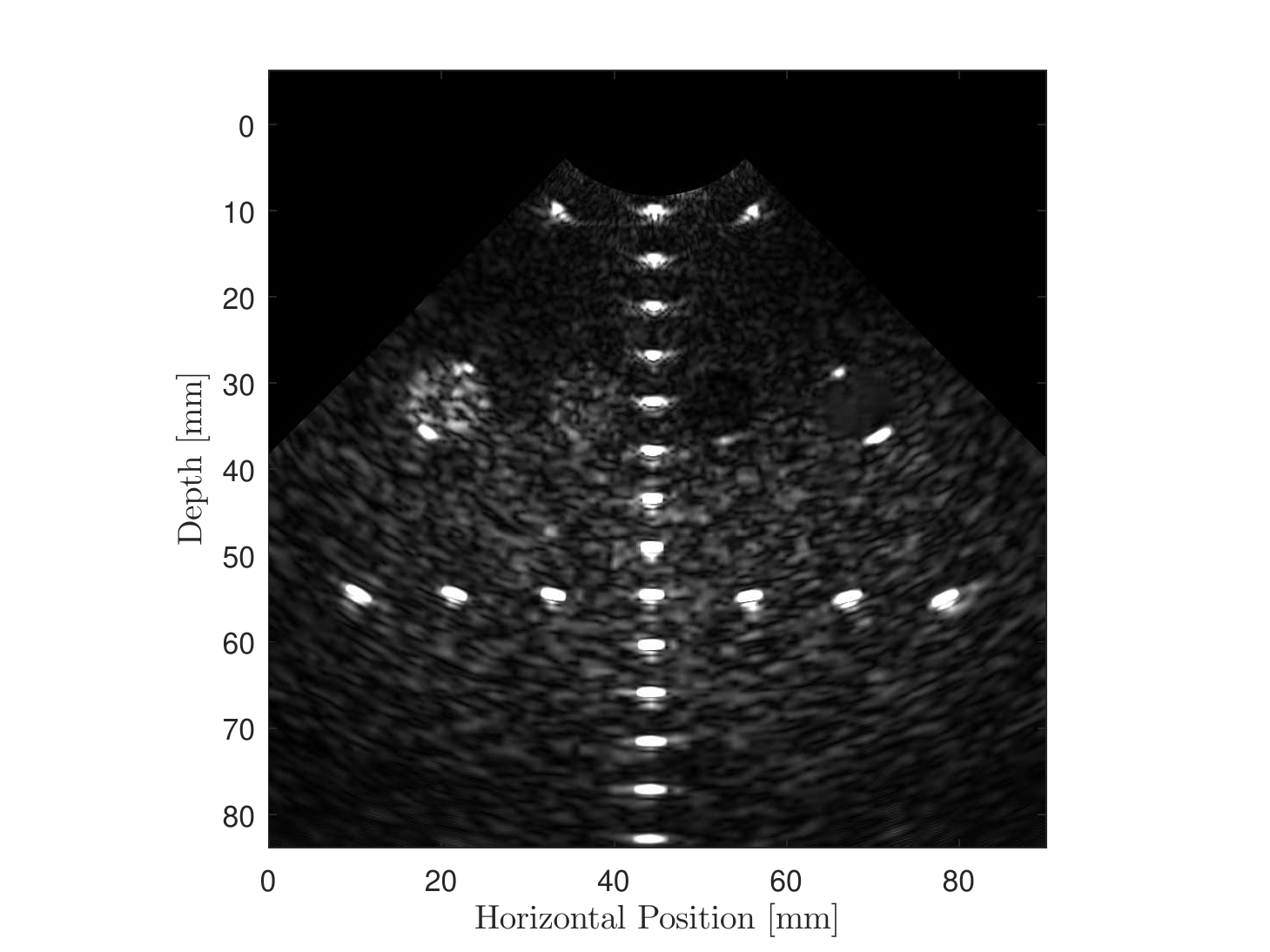}%
\label{fullImagePreComp}}\\
\centering
\subfloat[]{\includegraphics[width=9 cm]{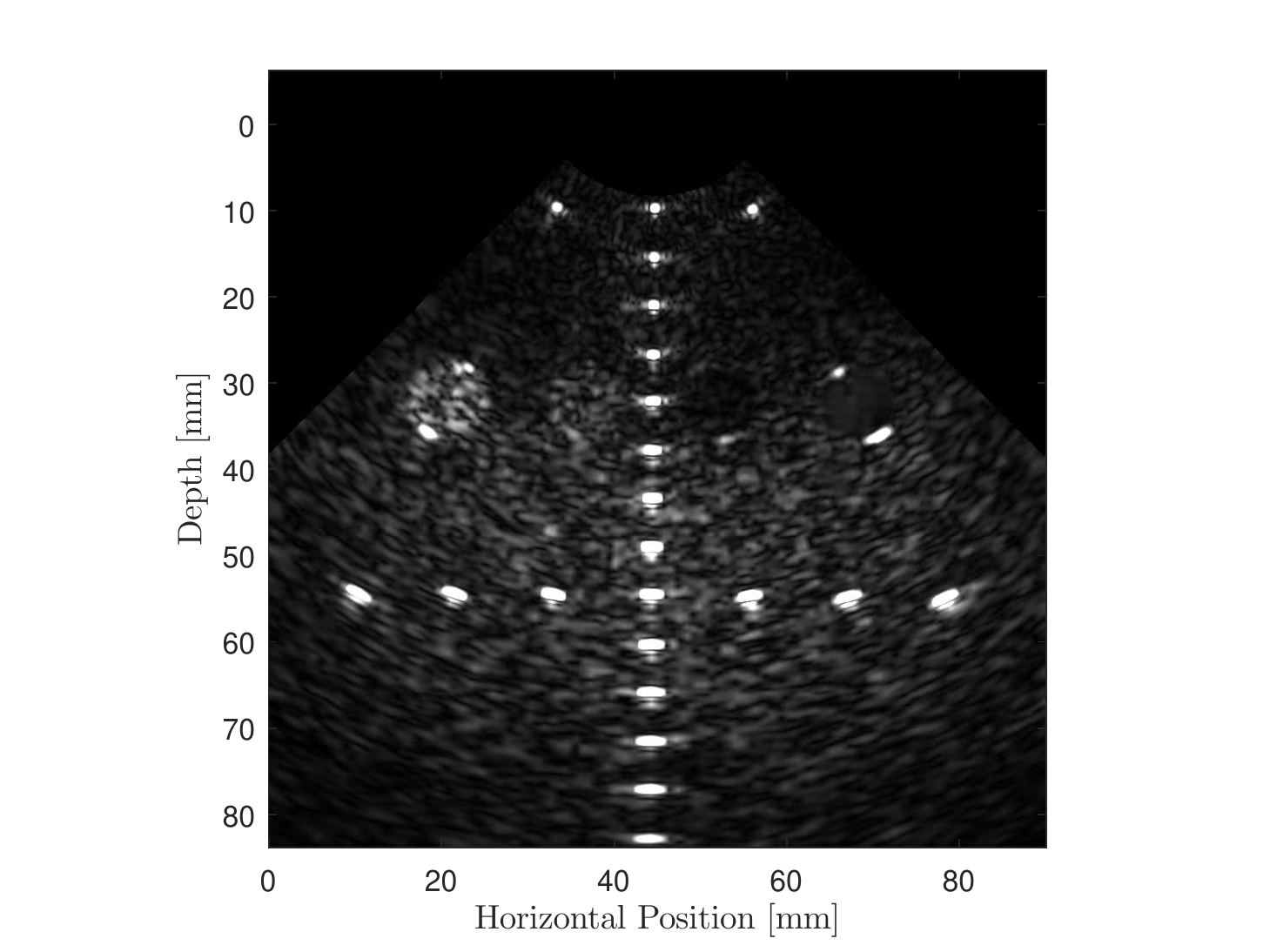}%
\label{fullImageFDBFQ29}}
\hfil
\subfloat[]{\includegraphics[width=9 cm]{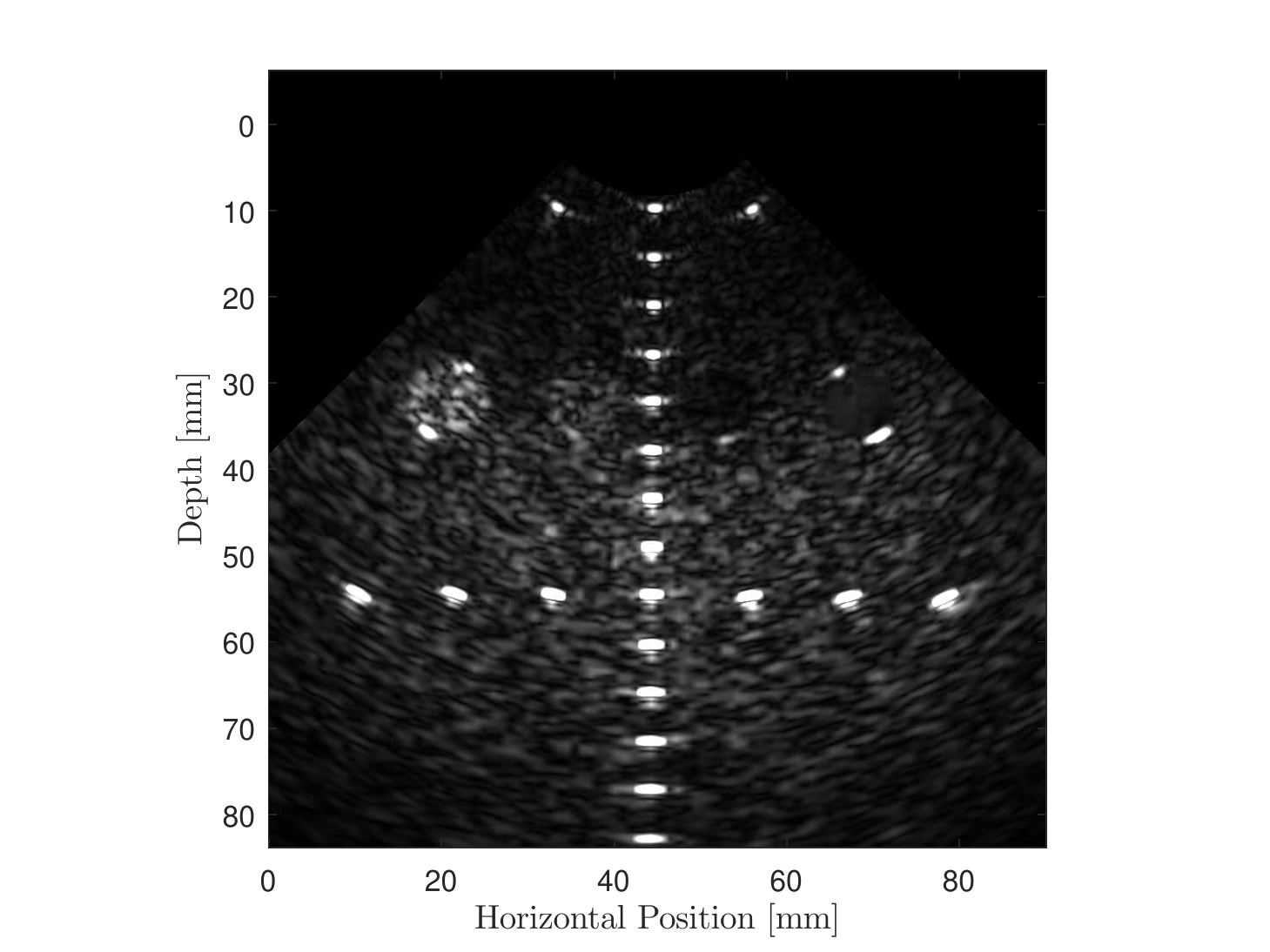}%
\label{fullImageFDBFQ9}}\\
\caption{Experimental results. Phantom scans obtained by: (a) time domain beamforming pre-compression, (b) time domain beamforming post-compression, (c) frequency domain beamforming with $N_q=29$, (d)  frequency domain beamforming with $N_q=9$.}.
\label{fig:fullImages}
\end{figure*}
\begin{figure*}[!t]
\centering
\subfloat[]{\includegraphics[width=9 cm]{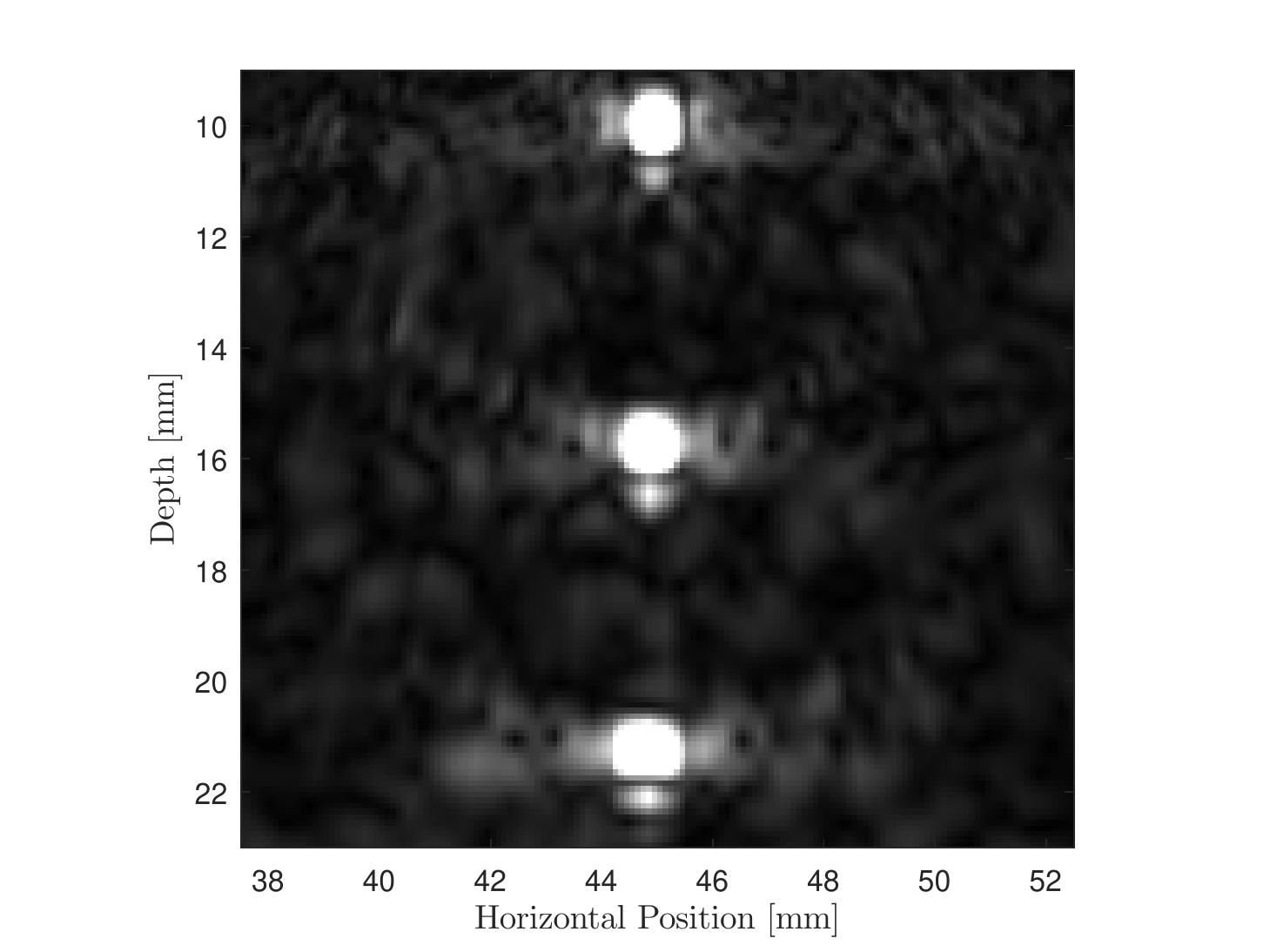}%
\label{fullImageZoomPostComp}}
\hfil
\subfloat[]{\includegraphics[width=9 cm]{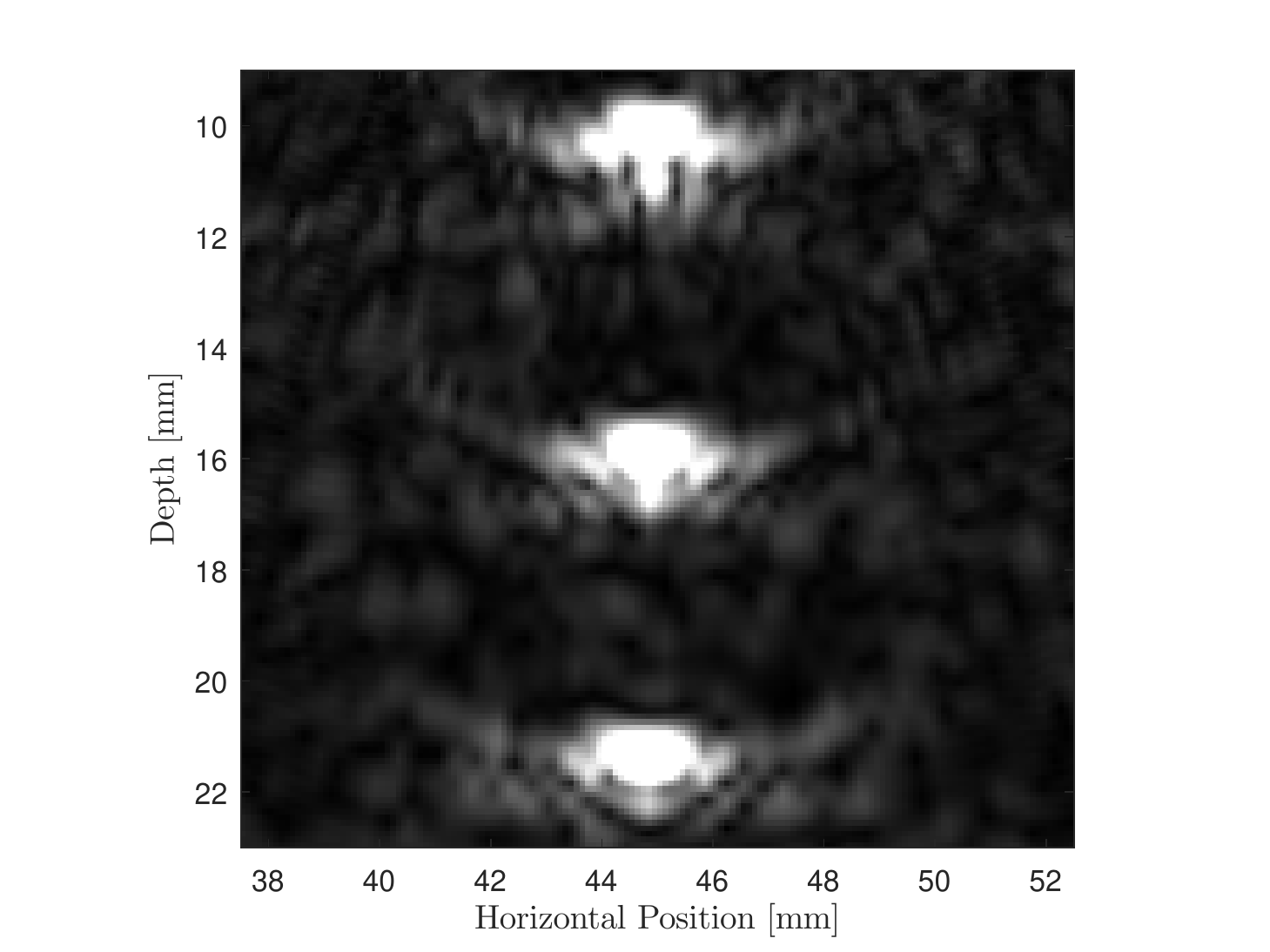}%
\label{fullImageZoomPreComp}}\\
\centering
\subfloat[]{\includegraphics[width=9 cm]{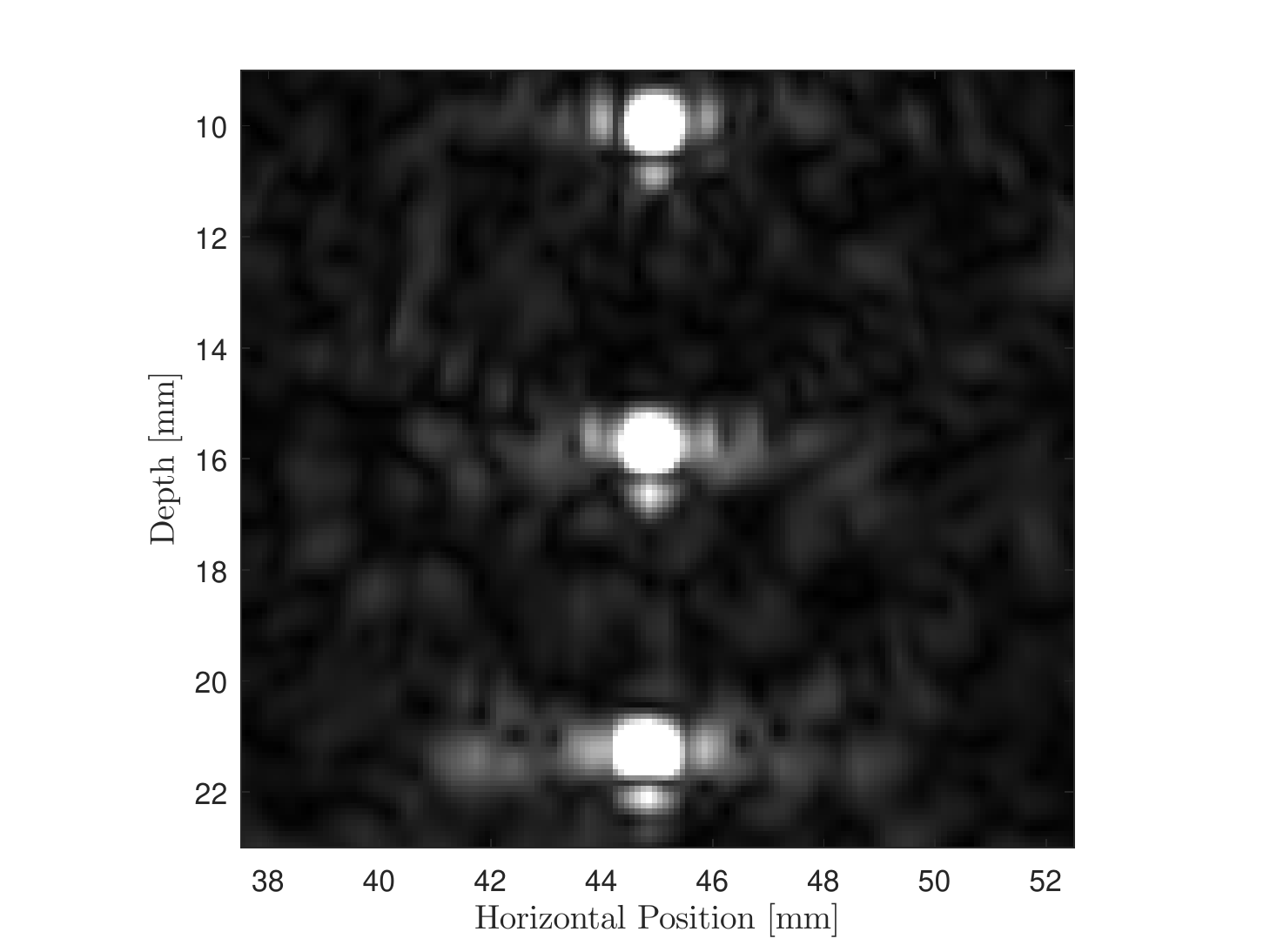}%
\label{fullImageZoomFDBFQ29}}
\hfil
\subfloat[]{\includegraphics[width=9 cm]{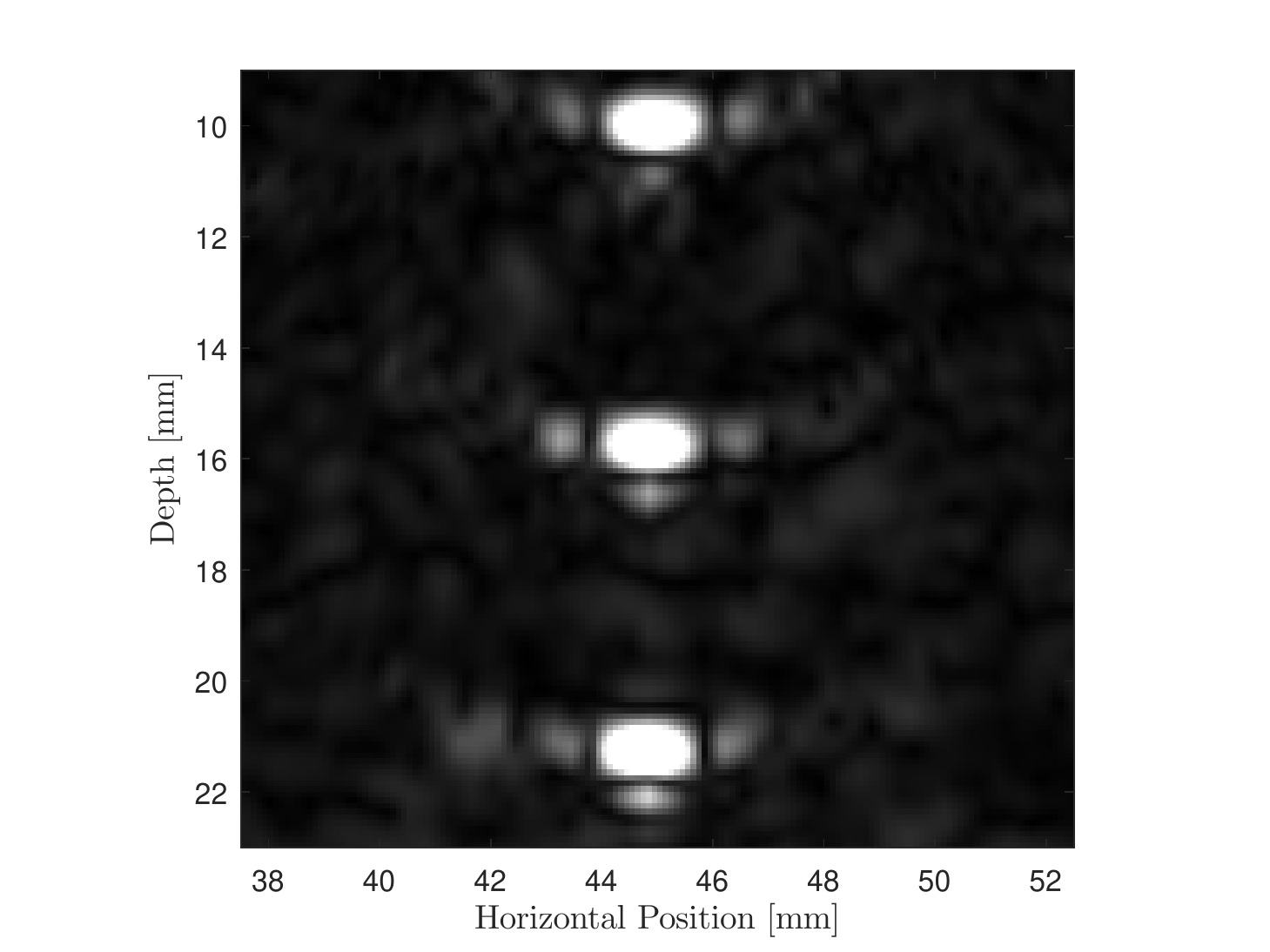}%
\label{fullImageZoomFDBFQ9}}\\
\caption{Experimental results, zoom in. Phantom scans obtained by: (a) time domain beamforming pre-compression, (b) time domain beamforming post-compression, (c) frequency domain beamforming with $N_q=29$, (d)  frequency domain beamforming with $N_q=9$.}.
\label{fig:fullImagesZoom}
\end{figure*}

\subsection{Experimental Setup}
In order to examine the performance of the proposed method we used real data acquired by Vantage 256, an ultrasound system of Verasonics. A tissue mimicking phantom Gammex 404GSLE of 90 mm depth was scanned by a 64-element phased array transducer P$4$-$2$v with a frequency response centered at $f_c = 2.9$ MHz. During acquisition, each element transmitted a linear FM defined in \eqref{eq:chirp} with time-bandwidth product $D=60$ and central frequency $f_0=3$ MHz. The raw data was processed using FoCUS defined in \eqref{eq:FourierDomainRelationship CE} for different approximation levels, namely different number of Q-coefficients, $N_q$. The performance in axial and lateral dimensions for different values of $N_q$ are compared to time domain beamforming pre- and post-compression. Saving in computational complexity for different values of $N_q$ are then discussed in Section \ref{ssec:comp complexity}.

\subsection{Imaging Results And Performance Analysis}
\label{ssec:Imaging results and performance}
We start with examination of the axial resolution of the proposed method. Figure \ref{fig:TDBF Vs FDBF single} presents scan-lines of point scatterer located at 10 mm from the transducer. As can be seen, the axial point spread functions of FoCUS and beamforming pre-compression are identical in terms of main lobe width, and their sidelobes are extremely close. This implies that FoCUS preserves the performance in terms of contrast and resolution in axial dimension. The above results are obtained for $N_q=9$, however, even coarser approximation corresponding to lower values of $N_q$ has almost no affect on the performance in the axial dimension. For $N_q=3$, the same main lobe width is obtained with only $1.3$ dB degradation in sidelobes level. This is significantly lower compared to 9 dB degradation of beamforimg post-compression.


The difference in the axial performance can be easily seen in the resulting images. Figure \ref{fullImagePostComp} and \ref{fullImagePreComp} show beamforming pre- and post-compression. FoCUS with $N_q=29$ and ${N_q=9}$ are presented in Figs. \ref{fullImageFDBFQ29} and \ref{fullImageFDBFQ9} respectively. A zoom in on the first three scatterers, presented in Fig. \ref{fig:fullImagesZoom}, verifies that the axial resolution of FoCUS with both $N_q=29$ and $N_q=9$ is the same as of beamforming pre-compression. In fact, it can be observed that Figs. \ref{fullImageFDBFQ29} and \ref{fullImagePostComp} look the same, implying that FoCUS achieves the same image quality as beamforming pre-compression which is optimal for coded imaging.



\begin{figure}
\begin{minipage}[b]{1.0\linewidth}
  \centering
  \centerline{\includegraphics[width=7cm]{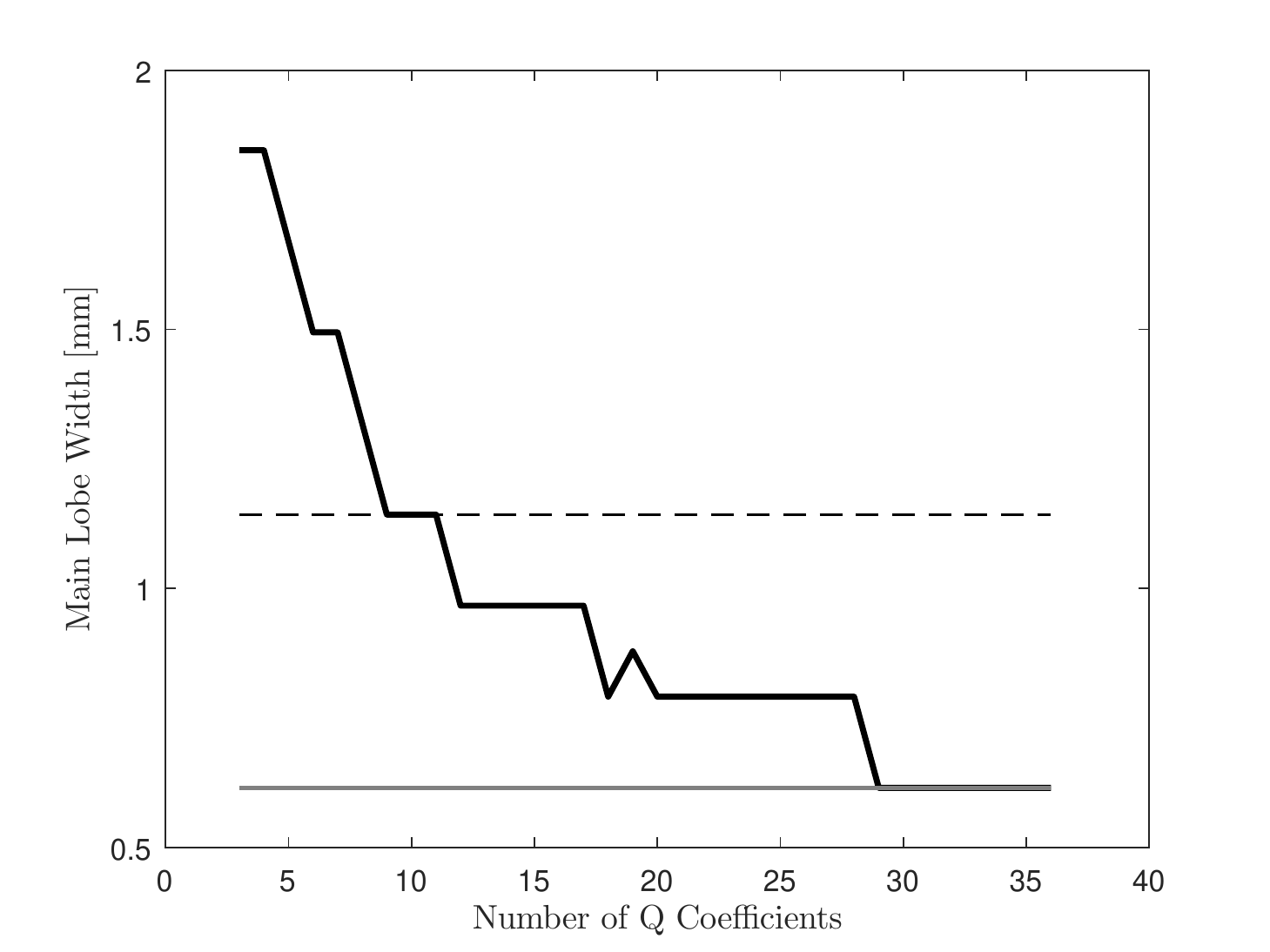}}
\end{minipage}
\caption{Main lobe width of the lateral PSF measured for a scatterer at 10 mm depth, as a function of $N_q$. Main lobe width obtained with beamforming pre- and post-compression are plotted in thin gray line and dashed line respectively.}
\label{fig:mainLobeVsNq}
\end{figure}

From Figs. \ref{fig:fullImages} and \ref{fig:fullImagesZoom} it is obvious that FoCUS outperforms beamforming post-compression. However the lateral resolution of FoCUS is decreased for $N_q=9$. As mentioned in Section \ref{sec:BF and PC in frequncy}, the number of Q-coefficients, controls the quality of the approximation presented in \eqref{eq:FourierDomainRelationship}. The results presented above show that this approximation affects mostly the performance in the lateral dimension. To study the effect of the approximation level on the lateral resolution, we measure the main lobe width of the lateral PSF corresponding to the scatterer located at 10 mm from the transducer. We compare it to beamforming post- and pre- compression. Figure \ref{fig:mainLobeVsNq} presents the main lobe width as a function of $N_q$. As expected, the lateral resolution decreases with $N_q$. For ${N_q}\geq29$, FoCUS obtains the same lateral resolution as beamforming pre-compression, and it outperforms beamforming post-compression as long as ${N_q}\geq9$.
Our results show that the proposed method obtains high image quality, and for $N_q$ large enough it yields the performance of beamforming pre-compression both in lateral and axial dimensions. The number of Q-coefficients, $N_q$, defines the reduction in computational complexity which will be discussed next.

\subsection{Computational Complexity}
\label{ssec:comp complexity}

For the derivation of the computational complexity of beamforming pre-compression and FoCUS we consider only multiplications. The number of samples comprising each scan-line is denoted by $N_s$, and is determined by the sampling rate and the imaging depth.
According to \eqref{eq:FourierDomainRelationship CE} the number of multiplications needed for a computation of one scan-line using $K$ coefficients from the set $\{c_{CE}[k]\}$ is:
\begin{equation}
\label{eq:CE complexity}
 N_{a}=MKN_q+\frac{N_s}{2}\log N_s,
\end{equation}
including the inverse Fourier transform. Here $M$ is the number of transducer elements and $N_q$ denotes the number of $\tilde{Q}_{k,m;\theta}[n]$ coefficients taken for the approximation in \eqref{eq:FourierDomainRelationship CE}.

When applying the conventional beamforming pre-compression, the computation includes the complexity of $M$ matched filters and interpolation of $M$ signals to apply the time-varying delays. Assuming linear complexity for the linear interpolation and an efficient MF implementation using FFT:
\begin{equation}
\label{eq:normal complexity}
 N_{b}=MN_s+M\left(\frac{3(N_s+N_h)}{2}\log(N_s+N_h)+N_s+N_h\right),
\end{equation}
multiplications are needed.




For sampling rate $f_s=4f_c$ the length of the MF is ${N_h=274}$ and $N_s=1392$. For a linear FM with time-bandwidth product $D=60$ and the above sampling rate, the bandwidth of the beamformed signal contains $K=260$ Fourier coefficients. Table \ref{table:sys_param} summarizes all the parameters.

\begin{figure}
\begin{minipage}[b]{1.0\linewidth}
  \centering
  \centerline{\includegraphics[width=7cm]{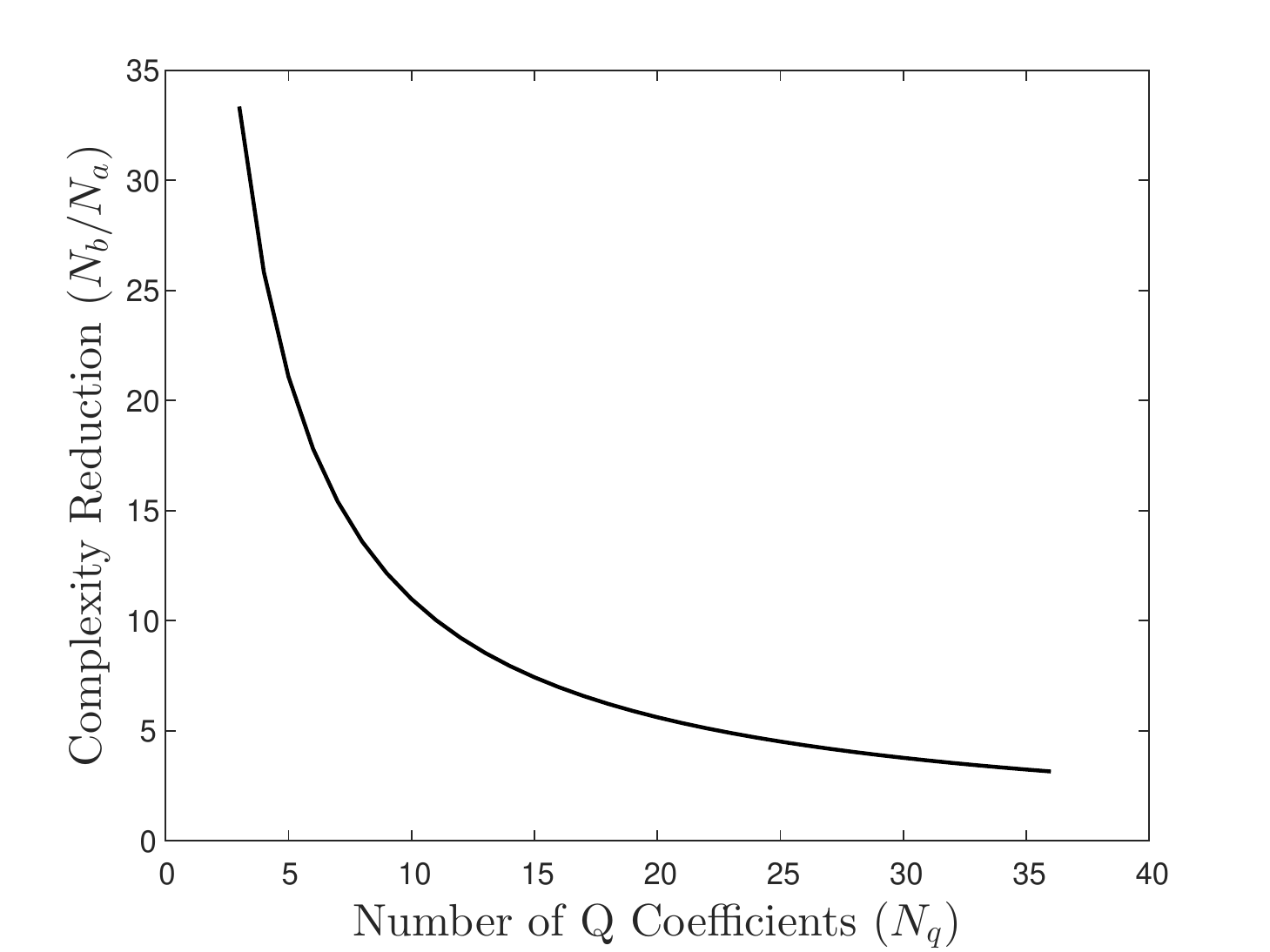}}
\end{minipage}
\caption{Computational complexity reduction as function of $N_q$.}
\label{fig:compReduction}
\end{figure}

\begin{table}[htdp] \caption{System parameters } \begin{center}
\vspace{0.5 cm}
\begin{tabular}{|l|l|c|}
  \hline
  \small{Parameter} & \small{Description} &  \small{Value}\\ \hline \hline
  \small{$f_c$} & \small{Transducer central frequency [MHz]} &\small{2.9} \\ \hline
  \small{$f_s$} & \small{Sampling frequency} &\small{$4f_c$} \\ \hline
  \small{$N_s$} & \small{Samples per element} &\small{1392} \\ \hline
  \small{$N_h$} & \small{Matched filter length} &\small{274} \\ \hline
  \small{K} & \small{Number of $c_{CE}[k]$ coefficients} &\small{260} \\ \hline
\end{tabular}
\end{center} \label{table:sys_param}
\end{table}
Figure \ref{fig:compReduction} shows the computational complexity reduction as a function of $N_q$. For $N_q=29$ our method achieves 4 fold complexity reduction while yielding optimal axial and lateral resolution. When choosing lower values for $N_q$ further complexity reduction is achieved yielding  up to 33 fold reduction for $N_q=3$. The lateral resolution is degraded accordingly as shown in Fig. \ref{fig:mainLobeVsNq}, while the axial resolution is preserved.

We note that $N_s$ and $N_h$ depend on the oversampling factor, $P=f_s/f_0$, and are given by
\begin{align}\label{eq: N_s and N_q}
  N_s &= T P f_c,  \\ \nonumber
  N_h &= D P.
\end{align}
Thus, $N_b/N_a$ increases with $P$, meaning that the saving in computational load is even more significant for higher oversampling factors. For example, taking $P=10$ leads to 11 fold reduction for $N_q=29$ and 77 fold reduction for $N_q=3$.


%

\section{Discussion and Conclusions}
\label{sec:discussion and conclusions}
In this paper we present a method allowing to avoid the extended computational load required by CE array imaging. The proposed approach is based on integration of pulse compression of each one of the detected signals to computationally efficient frequency domain processing. FDBF computes the Fourier coefficients of the beam as a weighted average of those of the detected signals. We show that in frequency, pulse compression of each channel can be performed together with beamforming by appropriate modification of the weights required for FDBF. As a result MF is applied at each detected signal without adding computational load to the FDBF technique.
Our method enables efficient implementation of CE in array imaging paving the way to enhanced SNR as well as improved imaging depth and frame-rate.

The proposed method, FoCUS, was implemented on a real system of Verasonics with a 64-element probe. We compared our results with beamforming pre-compression, which is the optimal implementation for CE array imaging. We evaluated the reduction in computational load for different oversampling factors. It was shown that for $P=4$ and $P=10$ FoCUS achieves the same image quality with 4 and 11 fold reduction in computational complexity respectively. Further reduction, up to 33 fold for $P=4$ and up to 77 for $P=10$, can be obtained at the expense of decreased lateral resolution. As a result our method enables adjusting lateral resolution to available computational power. It is important to mention that FoCUS preserves optimal axial resolution regardless of the reduction in complexity. This is in contrast to beamforming post-compression which is commonly used to reduced the amount of computations and severely degrades both axial and lateral resolution.

The performance in terms of lateral resolution can be potentially improved by post-processing coherence based techniques \cite{camacho2009phase,li2003adaptive}. The above methods are applied on beamformed data and result in up to two fold improvement of the lateral resolution. Integrating these methods with FoCUS may alow to obtain the maximal complexity reduction without compromising lateral resolution.

\bibliographystyle{IEEEtran}
\bibliography{IEEEfull,general}

\begin{thebibliography}{10}
\providecommand{\url}[1]{#1}
\csname url@samestyle\endcsname
\providecommand{\newblock}{\relax}
\providecommand{\bibinfo}[2]{#2}
\providecommand{\BIBentrySTDinterwordspacing}{\spaceskip=0pt\relax}
\providecommand{\BIBentryALTinterwordstretchfactor}{4}
\providecommand{\BIBentryALTinterwordspacing}{\spaceskip=\fontdimen2\font plus
\BIBentryALTinterwordstretchfactor\fontdimen3\font minus
  \fontdimen4\font\relax}
\providecommand{\BIBforeignlanguage}[2]{{%
\expandafter\ifx\csname l@#1\endcsname\relax
\typeout{** WARNING: IEEEtran.bst: No hyphenation pattern has been}%
\typeout{** loaded for the language `#1'. Using the pattern for}%
\typeout{** the default language instead.}%
\else
\language=\csname l@#1\endcsname
\fi
#2}}
\providecommand{\BIBdecl}{\relax}
\BIBdecl

\bibitem{rihaczek1969principles}
A.~W. Rihaczek, \emph{Principles of high-resolution radar}.\hskip 1em plus
  0.5em minus 0.4em\relax McGraw-Hill New York, 1969.

\bibitem{chiao2005coded}
R.~Y. Chiao and X.~Hao, ``Coded excitation for diagnostic ultrasound: a system
  developer's perspective,'' \emph{IEEE Transactions on Ultrasonics,
  Ferroelectrics, and Frequency Control}, vol.~2, no.~52, pp. 160--170, 2005.

\bibitem{misaridis2005use2}
T.~Misaridis and J.~A. Jensen, ``Use of modulated excitation signals in medical
  ultrasound. \textsc{P}art \textsc{II}: Design and performance for medical
  imaging applications,'' \emph{IEEE Transactions on Ultrasonics,
  Ferroelectrics and Frequency Control}, vol.~52, no.~2, pp. 192--207, 2005.

\bibitem{o1992coded}
M.~O'Donnell, ``Coded excitation system for improving the penetration of
  real-time phased-array imaging systems,'' \emph{IEEE transactions on
  ultrasonics, ferroelectrics, and frequency control}, vol.~39, no.~3, pp.
  341--351, 1992.

\bibitem{liu2005coded}
J.~Liu and M.~F. Insana, ``Coded pulse excitation for ultrasonic strain
  imaging,'' \emph{IEEE transactions on ultrasonics, ferroelectrics, and
  frequency control}, vol.~52, no.~2, pp. 231--240, 2005.

\bibitem{vogt2005development}
M.~Vogt and H.~Ermert, ``Development and evaluation of a high-frequency
  ultrasound-based system for in vivo strain imaging of the skin,'' \emph{IEEE
  transactions on ultrasonics, ferroelectrics, and frequency control}, vol.~52,
  no.~3, pp. 375--385, 2005.

\bibitem{novell2009contrast}
A.~Novell, S.~Van Der~Meer, M.~Versluis, N.~De~Jong, and A.~Bouakaz, ``Contrast
  agent response to chirp reversal: simulations, optical observations, and
  acoustical verification,'' \emph{IEEE transactions on ultrasonics,
  ferroelectrics, and frequency control}, vol.~56, no.~6, pp. 1199--1206, 2009.

\bibitem{muzilla1999method}
D.~J. Muzilla, R.~Y. Chiao, and A.~L. Hall, ``Method and apparatus for color
  flow imaging using coded excitation with single codes,'' Aug.~17 1999, uS
  Patent 5,938,611.

\bibitem{gammelmark2003multielement}
K.~L. Gammelmark and J.~A. Jensen, ``Multielement synthetic transmit aperture
  imaging using temporal encoding,'' \emph{IEEE transactions on medical
  imaging}, vol.~22, no.~4, pp. 552--563, 2003.

\bibitem{o2005coded}
M.~O'Donnell and Y.~Wang, ``Coded excitation for synthetic aperture ultrasound
  imaging,'' \emph{IEEE transactions on ultrasonics, ferroelectrics, and
  frequency control}, vol.~52, no.~2, pp. 171--176, 2005.

\bibitem{song2015coded}
P.~Song, M.~W. Urban, A.~Manduca, J.~F. Greenleaf, and S.~Chen, ``Coded
  excitation plane wave imaging for shear wave motion detection,'' \emph{IEEE
  transactions on ultrasonics, ferroelectrics, and frequency control}, vol.~62,
  no.~7, pp. 1356--1372, 2015.

\bibitem{lewandowski2008high}
M.~Lewandowski and A.~Nowicki, ``High frequency coded imaging system with rf,''
  \emph{IEEE transactions on ultrasonics, ferroelectrics, and frequency
  control}, vol.~55, no.~8, pp. 1878--1882, 2008.

\bibitem{misaridis2005use1}
T.~Misaridis and J.~A. Jensen, ``Use of modulated excitation signals in medical
  ultrasound. \textsc{P}art \textsc{I}: Basic concepts and expected benefits,''
  \emph{IEEE Transactions on Ultrasonics, Ferroelectrics and Frequency
  Control}, vol.~52, no.~2, pp. 177--191, 2005.

\bibitem{misaridis2005use3}
------, ``Use of modulated excitation signals in medical ultrasound.
  \textsc{P}art \textsc{III}: high frame rate imaging,'' \emph{IEEE
  Transactions on Ultrasonics, Ferroelectrics and Frequency Control}, vol.~52,
  no.~2, pp. 208--219, 2005.

\bibitem{bjerngaard2002should}
R.~Bjerngaard and J.~A. Jensen, ``Should compression of coded waveforms be done
  before or after focusing?'' in \emph{Medical Imaging 2002}.\hskip 1em plus
  0.5em minus 0.4em\relax International Society for Optics and Photonics, 2002,
  pp. 47--58.

\bibitem{yoon2013efficient}
C.~Yoon, W.~Lee, J.~H. Chang, T.~Song, and Y.~Yoo, ``An efficient pulse
  compression method of chirp-coded excitation in medical ultrasound imaging,''
  \emph{IEEE transactions on ultrasonics, ferroelectrics, and frequency
  control}, vol.~60, no.~10, pp. 2225--2229, 2013.

\bibitem{ramalli2015real}
A.~Ramalli, F.~Guidi, E.~Boni, and P.~Tortoli, ``A real-time chirp-coded
  imaging system with tissue attenuation compensation,'' \emph{Ultrasonics},
  vol.~60, pp. 65--75, 2015.

\bibitem{steinberg1992digital}
B.~D. Steinberg, ``Digital beamforming in ultrasound,'' \emph{IEEE Transactions
  on Ultrasonics, Ferroelectrics and Frequency Control}, vol.~39, no.~6, pp.
  716--721, 1992.

\bibitem{wagner2012compressed}
N.~Wagner, Y.~C. Eldar, and Z.~Friedman, ``Compressed beamforming in ultrasound
  imaging,'' \emph{IEEE Transactions on Signal Processing}, vol.~60, no.~9, pp.
  4643--4657, 2012.

\bibitem{chernyakova2014compressed}
T.~Chernyakova and Y.~C. Eldar, ``Fourier domain beamforming: The path to
  compressed ultrasound imaging,'' \emph{IEEE Transactions on Ultrasonics,
  Ferroelectronics, and Frequency Control}, vol.~61, no.~8, pp. 1252--1267,
  2014.

\bibitem{burshtein2016sub}
A.~Burshtein, M.~Birk, T.~Chernyakova, A.~Eilam, A.~Kempinski, and Y.~C. Eldar,
  ``Sub-nyquist sampling and fourier domain beamforming in volumetric
  ultrasound imaging,'' \emph{IEEE transactions on ultrasonics, ferroelectrics,
  and frequency control}, vol.~63, no.~5, pp. 703--716, 2016.

\bibitem{eldar2015sampling}
Y.~C. Eldar, \emph{Sampling Theory: Beyond Bandlimited Systems}.\hskip 1em plus
  0.5em minus 0.4em\relax Cambridge University Press, 2015.

\bibitem{wehner1995high}
D.~R. Wehner, ``High resolution radar (2nd),'' \emph{Edition. Artech House
  Inc}, 1995.

\bibitem{jensen1996estimation}
J.~A. Jensen, \emph{Estimation of blood velocities using ultrasound: a signal
  processing approach}.\hskip 1em plus 0.5em minus 0.4em\relax Cambridge
  University Press, 1996.

\bibitem{cook1967radar}
C.~Cook and M.~Bernfeld, ``Radar signals—an introduction to theory and
  practice,'' 1967.

\bibitem{van2004detection}
H.~L. Van~Trees, \emph{Detection, Estimation, and Modulation Theory, Optimum
  Array Processing}.\hskip 1em plus 0.5em minus 0.4em\relax Wiley-Interscience,
  2004.

\bibitem{williams1968fast}
J.~R. Williams, ``Fast beam-forming algorithm,'' \emph{The Journal of the
  Acoustical Society of America}, vol.~44, p. 1454, 1968.

\bibitem{rudnick1969digital}
P.~Rudnick, ``Digital beamforming in the frequency domain,'' \emph{The Journal
  of the Acoustical Society of America}, vol.~46, p. 1089, 1969.

\bibitem{camacho2009phase}
J.~Camacho, M.~Parrilla, and C.~Fritsch, ``Phase coherence imaging.''
  \emph{IEEE transactions on ultrasonics, ferroelectrics, and frequency
  control}, vol.~56, no.~5, pp. 958--974, 2009.

\bibitem{li2003adaptive}
P.~C. Li and M.~L. Li, ``Adaptive imaging using the generalized coherence
  factor,'' \emph{IEEE transactions on ultrasonics, ferroelectrics, and
  frequency control}, vol.~50, no.~2, pp. 128--141, 2003.

\end{thebibliography}

\end{document}